\documentstyle[aas2pp4]{article}

\begin{document}

\def\simlt{$_<\atop{^\sim}$}
\def\simgt{$_>\atop{^\sim}$}
\def\q0{q$_0$}
\def\h0{H$_0$}
\def\Wg{$\Omega_g$}
\def\Wlamb{$\Omega_{\rm\Lambda}$}
\def\Wmatter{$\Omega_{\rm M}$}
\def\Wgz{$\Omega_g(z)$}
\def\Wstar{$\Omega_{\ast}$}
\def\Rgstar{R$_{g\ast}$}
\def\etal{{\it et~al.}}
\def\eg{e.g.~}
\def\ie{i.e.~}
\def\cf{c.f.~}
\def\lya{Ly$\alpha$\ }
\def\kms{km~s$^{-1}$ }
\def\kmsM{km~s$^{-1}$~Mpc}
\def\cm2{cm$^{-2}$}
\def\nhi{\mbox{$N_{\rm HI}$}}
\def\atomcm{atoms cm$^{-2}$}

%\slugcomment{}

%\shorttitle{Storrie-Lombardi \& Wolfe}
%\shortauthors{Surveys for z $>$ 3 Damped \lya Absorption Systems}

\title{Surveys for z $>$ 3 Damped \lya Absorption Systems: \\ the Evolution
of Neutral Gas\footnotemark[1]}

\author{ Lisa J. Storrie-Lombardi} 
\affil{SIRTF Science Center, Caltech, MS 100-22, Pasadena, CA 91125, lisa@ipac.caltech.edu}
%\email{lisa@ipac.caltech.edu}

\and

\author{ Arthur M. Wolfe}
\affil{Department of Physics, and Center for Astrophysics and Space Sciences \\
University of California--San Diego,
La Jolla, CA 92093-0424}

\author{Accepted for publication in the Astrophysical Journal}

\footnotetext[1]{
Some of the observations presented
here were obtained at the W.M.~Keck Observatory, which is
operated as a scientific partnership among the California Institute of
Technology, the University of California and the National Aeronautics and
Space Administration. The Observatory was made possible by the generous
financial support of the W.M.~Keck Foundation.  Observations were
also made using the 3.9-m Anglo-Australian Telescope at the
Anglo-Australian Observatory and the Shane 3-m Telescope
at the Lick Observatory.}

\begin{abstract} 
We have completed spectroscopic observations using LRIS on the
Keck 1 telescope of 30 very high redshift quasars, 11 selected
for the presence of damped \lya absorption systems and 
19 with redshifts z $>$ 3.5 not previously surveyed for 
absorption systems. We have surveyed an additional 
10 QSOs with the Lick 120'' and the Anglo-Australian Telescope.
We have combined these with previous data resulting
in a statistical sample of 646 QSOs and 85 damped \lya 
absorbers with column densities {\nhi} $\ge 2 \times 10^{20}$ 
{\atomcm} covering the redshift range 0.008 $\le$ z $\le$ 4.694.
Four main features of how the neutral gas in the Universe
evolves with redshift are evident from these data. 

\begin{enumerate}
\item  For the first time we determine a {\it statistically significant}
steepening in the column density distribution 
function at redshifts z $>$ 4.0 ($> 99.7\%$ confidence).
The steepening of the distribution function is 
due to both fewer very high column density absorbers
(\nhi $\ge$ $10^{21}$ \atomcm) and more lower column density 
systems (\nhi $= 2-4 \times 10^{20}$ \atomcm).

\item  The frequency of very high column
density absorbers (\nhi $\ge$ $10^{21}$ \atomcm) reaches
a peak in the redshift range $1.5 < z < 4$ 
when the Universe is 10-30\% of its present age.
Though the sample size is still small the peak 
epoch appears to be 3.0 $\le$ z $\le$ 3.5.
The highest column density absorbers disappear
rapidly towards higher redshifts in the range
$z=3.5 \rightarrow 4.7$ and lower redshifts
$z=3.0 \rightarrow 0$.
None with column densities log \nhi $\ge$ 21 have yet been detected
at $z > 4$ though we have increased the redshift path surveyed
$\approx$ 60\%. 

\item With our current data set, the comoving mass density of neutral gas,
\Wg, appears to peak at $ 3.0 < z < 3.5$, but the uncertainties
are still too large to determine the precise shape of the 
{\Wg}.  The statistics 
are consistent with a constant value of {\Wg} for $2 < z < 4$.
There is still tentative evidence for a dropoff at $z > 4$ as indicated by
earlier data sets.  
If we define \Rgstar $\equiv$ \Wg$/$\Wstar,
where {\Rgstar} is the ratio of the 
peak value of {\Wg} to  
\Wstar, the mass density in galaxies in the local universe,
we find values 
of {\Rgstar} = 0.25 - 0.5 at $z \sim 3$ depending on the cosmology.  
For an $\Omega = 1$ Universe with a zero cosmological constant 
{\Rgstar} $\approx$ 0.5.  
For an $\Omega = 1$ Universe with a positive 
cosmological constant (\Wlamb = 0.7, \Wmatter = 0.3) we find 
{\Rgstar} $\approx$ 0.25.  
For a universe with \Wlamb = 0 and \Wmatter = 0.3,  
the value of {\Rgstar} $\approx$ 0.3. 

\item {\Wg} decreases with redshift for the interval $z=3.5 \rightarrow 0.008$
for our data set, but we briefly discuss new results from Rao and 
Turnshek for $z < 1.5$ that suggest \Wg (z$<$1.5) may be higher than 
previously determined. 

\end{enumerate}

To make the data in our statistical sample more readily available
for comparison with scenarios from various cosmological models, 
we provide tables that includes all 646 QSOs from our new survey 
and previously published surveys. 
They list the minimum and maximum redshift defining the 
redshift path along each line of sight, the QSO emission redshift,
and when an absorber is detected, 
the absorption redshift and measured HI column density. 

\end{abstract}

\section{Introduction}

Surveys for damped \lya absorption systems have been 
driven historically by the desire to detect galaxies
in very early stages of evolution, before most of the
gas has turned in to stars. 
Because the gas should be neutral and at high column densities 
({\nhi} $>$ $10^{20}$ \atomcm) it will leave spectroscopic imprints on the light
emitted by background QSOs. These absorbers are identified as damped \lya 
absorption systems. 
Previous studies of the evolution of the neutral gas
(Wolfe \etal\ 1986; Lanzetta \etal 1991; Wolfe \etal  1995 [WLFC95]; 
Storrie-Lombardi, McMahon \& Irwin 1996 [SMI96]),
their metal abundances (Lu \etal\  1996; Pettini \etal\ 1997; Prochaska
\& Wolfe 1999),
their dust content (Fall \& Pei 1993; Pei \& Fall 1995; Pei, Fall \& Hauser 1999), 
and their kinematics (Prochaska \& Wolfe 1997, 1998a, 1998b) 
provide compelling evidence that damped absorbers are the progenitors 
of present day galaxies.  However, substantial debate continues
over exactly which galaxies these are (\eg\  Le Brun \etal\ 1997; 
Haehnelt, Steinmetz \& Rauch 1998;
Pettini \etal\ 1999; Prochaska \& Wolfe 1998a, 1998b; Rao \& Turnshek 1998;
Salucci \& Persic 1999).
We present results of a new survey for damped \lya absorbers in the
redshift range $z$ $\approx$ 3 $-$ 4.5.

The comoving mass density of neutral gas, \Wgz,  
at $z$ $\approx$ 3 has been known for some time to 
be comparable to the density of visible matter, \ie\ stars
in present day galaxies for an $\Omega = 1$ ($\Lambda = 0$) universe 
(Rao \& Briggs 1993; WLFC95; SMI96). 
We have improved our estimates of \Wgz\ at high redshift 
with new surveys (this work; Storrie-Lombardi \& Hook 2000) and 
estimates of {\Wstar} have also been updated 
(Fukugita, Hogan \& Peebles 1998).
The first damped \lya survey to have a substantial 
data set with redshifts z $>$ 3.0 (Storrie-Lombardi \etal\ 1996 [SMIH96]) 
hinted at a  turnover in \Wgz\ at redshifts z $>$ 4 (SMI96), prior
to which damped \lya alpha systems might still be collapsing.
In this paper we update the discussion of these issues  
with a larger high redshift data set.

All of the follow-up observations of known candidate 
damped \lya absorbers were carried out with the 
Low Resolution Imaging Spectrometer (LRIS; Oke \etal\ 1995)
on the Keck 1 telescope.
The signal-to-noise ratios of the resulting spectra
are $\approx$  25:1 per pixel and the resolution is $\approx$
2 {\AA} thus assuring the spectra are adequate for testing the
damping hypothesis (cf. WLFC95).
The survey data for new damped \lya candidates was carried
out with LRIS on Keck, the KAST spectrograph at the Lick 120'', and 
the RGO spectrograph at the Anglo-Australian Telescope.

The paper is organized as follows. 
In $\S$2 we present the spectra along with fits of 
Voigt profiles to the damped \lya lines to 
accurately determine the column densities. 
$\S$3 contains a compilation of a statistical sample
of damped \lya systems including results from all 
available previous surveys.
In $\S$4 we present analyses of the 3 statistical
quantities characterizing the evolution of the damped systems,
namely, the number of systems per unit redshift interval,
$dn/dz$, the comoving mass density of neutral gas in units
of the current critical density, \Wgz, and 
the frequency distribution of HI column
densities, $f(N,z)$. 
%We compare the number of sightlines with
%single and double damped sytems to place constraints
%on the effects of gravitational lensing on $dn/dz$.
We discuss our results in $\S$5.

\section{Data}

In this section we first present follow-up spectra of previously
reported damped \lya candidates and then discuss new candidates
discovered in our survey.

\subsection{LRIS Spectroscopy}

Table~\ref{t1} is a journal of observations describing the LRIS
observations of 30 QSOs. 
Column 1 gives the QSO coordinate
name, column 2 the emission redshift, 
column 3 the UT date of the observations,
columns 4 and 5 the $V$ and/or $R$ magnitude of the QSO, column 6 the
central wavelength of the grating setting, column 7 the exposure time,
column 8 the number of grooves per mm of the grating, and column 
9 a reference to observations reporting discovery of the candidate.
The spectra were
wavelength calibrated with a ThAr lamp, flat fielded with a quartz
lamp, and reduced in the usual manner.
 
\subsubsection{Objects Selected for the Presence of Damped \lya Candidates}
\label{s_0953}

Table~\ref{t2} presents results for the subset of 11 QSOs selected
for the presence of damped
\lya candidates. We include only those candidates found in
surveys specifically designed to search
for strong absorption features blueward of \lya emission.
The confirmed damped \lya systems are thus included
in our statistical sample discussed in $\S$ 4 that covers
lines of sight observed with sufficient accuracy that
damped absorption lines arising
in gas with {\nhi} above the threshold, 2$\times 10^{20}$ \atomcm, could have
been detected at the 5$\sigma$ confidence level. 
Column 1 gives the QSO coordinate
name, column 2 the emission redshift,
columns 3 and 4 the $V$ or $R$ magnitude of the
QSO, columns 5$-$8 give the reference to
the low resolution survey that identified the candidate absorber, the absorption
redshift, the \lya equivalent width, and the inferred
log $N$(HI). Columns 9 and 10
give the absorption redshift and log $N$(HI) 
deduced by profile fitting from the
higher resolution LRIS spectra.

Figure~\ref{f_specconf} exhibits
damped \lya absorption spectra for
the 10 confirmed damped \lya systems with log {\nhi} $\ge$ 20.3
cm$^{-2}$ and the $z$ = 2.8228 system toward Q0249$-$2212 in which
log {\nhi} = 20.20 cm$^{-2}$.
Three Voigt profile fits are shown
in order of increasing {\nhi} with the middle
value representing the mean {\nhi} inferred from the optimal fit,
and the other two judged to be $\pm$1 $sigma$ from the mean. The results
of the fits are given in table~\ref{t2}.

Because the high density \lya forest in this redshift
range can lead to ambiguities
in the fitting procedure we did a number of simulations 
creating spectra with damped absorbers in the forest at 
known redshifts and column densities, and testing our fitting
techniques on these prior to making the final fits on the 
object spectra. These are extensions of the simulations discussed
in SMIH96 where the original damped 
candidates were selected. 
The continua of the spectra were first fit
with an automated task using cubic splines. These were then
modified interactively in dense forest regions where the 
continuum was placed too low, i.e. the fit had dipped into
the line.  In every case the redshift of the fit
is fixed by the low-ion metal lines
detected redward of \lya emission; \ie, outside
the \lya forest, and a chi square minimization
technique, {\bf vpgti} \footnote{Written by R.F. Carswell, J.K. Webb, 
A.J. Cooke, and M.J. Irwin.}
is subsequently used to make the fit. 

The metal lines were taken from the LRIS spectra when we 
had wavelength coverage that went far enough into the red to 
include them. In the other cases we used the metal line
redshifts determined from 5\AA\ resolution spectra (SMIH96)
taken with the ISIS spectrograph
on the William Herschel Telescope (WHT). We confirmed
that the wavelength calibrations for the WHT and LRIS data
agree well in QSOs with overlap regions. 
The maximum difference in the redshift we determined from
the same line measured in different spectra is 0.002
so we are confident that the metal lines in the WHT spectra
don't introduce any significant error when applying the measured
redshift to the LRIS spectra.

As expected, the higher resolution of the LRIS 
spectra make the fitting less ambiguous than the candidate 
selection process, though the agreement
between the estimated and actual column density is very good
in almost all the cases. It was crucial to have a low ionization 
metal line to anchor the redshift for the HI absorption, and 
in all cases one or both wings of the damped feature provided a good
constraint on the column density measurement.  Weaker HI 
components were also used to help constrain the fits, but only
the final damped components are shown in the figures.

The following are brief comments on the individual fits:

\noindent (1) BR B0019$-$1522, $z_{abs}$ = 3.4370, log {\nhi}=20.92

Si II 1527, Fe II 1608, Al II 1670  were used
to determine the redshift. 
This is the only damped system
where the column density estimated from the 5\AA\ resolution
WHT spectrum was substantially in error. This is due to the fact
that it lies at 5400\AA\, which is precisely where the blue 
and red arms of the ISIS spectra were pasted together. 

\noindent (2) BRI B0111$-$2819, $z_{abs}$ = 3.1043, log {\nhi}=21.0

Si II 1808 was used to determine the redshift.
We did not use the Fe II 1608 and Al II 1670 transitions because 
they were very broad. We also fitted a weaker component 
with $z$ = 3.070 and log {\nhi} = 19.5 cm$^{-2}$.

\noindent (3) Q0249$-$2212, $z_{abs}$ = 2.8228, log {\nhi}=20.2

Al II 1670 was used to fix the redshift. The
fit is excellent owing to absence of \lya forest
confusion noise near the center of the line profile.
Note that the fitted column density, log {\nhi} = 20.2
cm$^{-2}$ removes this object from the statistical sample.

\noindent (4) BR B0951$-$0450, $z_{abs}$ = 3.8580, log {\nhi}=20.6

In common with many other damped
\lya systems  observed at $z$ $>$ 3,
substantial \lya forest structure interferes with the
fitting procedure in the wings of the damped feature. 
In this case we used Si II 1527 observed at low resolution
(FWHM = 5 {\AA}) with the WHT to anchor the redshift.
The column density is constrained by the bump at 5875\AA\ and 
the wings on both sides.

\noindent (5) BR B0951$-$0450, $z_{abs}$ = 4.2028, log {\nhi}=20.4
    
We again used Si II 1527 from the WHT spectrum to determine  
the redshift. The fit is very good due to
the lack of \lya forest confusion near the line center.

\noindent (6) BRI B0952$-$0115, $z_{abs}$ = 4.0238, log {\nhi}=20.55

A strong C II 1334 line from the LRIS
spectrum anchors the fit. The figure shows that the
fit is somewhat uncertain owing to strong \lya absorption
in the blue wing of the damped profile. However, the resulting
log {\nhi} is well determined because the red wing and the
blue positive feature near 6085 {\AA} place strong constraints
on the fit.

\noindent (7) BRI B1013$+$0035, $z_{abs}$ = 3.1031, log {\nhi}=21.1

We used metal lines from the WHT spectrum 
to fix the redshift because our limited
coverage with LRIS placed all the strong low-ion metal
transitions in the \lya forest. The \lya forest
confusion is especially strong in the blue wing of the
damping profile. Nevertheless the fit is again constrained
by the observed red wing of the profile and the postive
spikes blueward of the profile.

\noindent (8) BRI B1114$-$0922, $z_{abs}$ = 4.2576, log {\nhi}=20.3

There is no LRIS coverage redward of \lya emission,
so we again used metal lines from the WHT spectrum.
However, the fit was well constrained by statistically
significant positive spikes on both red and blue wings
of the damping profile.

\noindent (9) BRI B1346$-$0322, $z_{abs}$ = 3.7343, log {\nhi}=20.72

This damping profile is the best determined of this sample
because the redshift is fixed by a single strong C II line in the LRIS spectrum 
In addition the absence of significant \lya forest absorption in
the profile wings results in an excellent fit.

\noindent (10) BRI B1500$+$0824, $z_{abs}$ = 2.7968, log {\nhi}=20.8

Because of limited integration time, this is the
noisiest spectrum we obtained. However, the rise of the
spectrum to the red of the damped \lya trough places
an upper limit not far above the quoted value for
{\nhi}. The greater uncertainties are reflected in the
larger error bars of the fit. We used WHT metal lines to
anchor the redshift, because of our limited
LRIS coverage.

\noindent (11) BR B2237$-$0607, $z_{abs}$ = 4.0691, log {\nhi}=20.5

In this case we used C II 1334 which had 2 strong absorption
components. Both were used to form the fit
shown in figure~\ref{f_specconf}. It is well constrained owing to
the high signal-to-noise ratio of the data.
This object has since been observed with HIRES (Lu \etal\ 1996)
and their result agrees well with ours (log {\nhi} = 20.5).

\subsubsection{New LRIS Survey for Damped Ly$\alpha$ Systems }

We obtained LRIS spectra for 19 high redshift QSOs that
had not been previously studied for the presence of damped \lya lines
and one QSO with a known damped absorber (Q0201$+$1120).
The objects not previously surveyed come from several quasar samples 
(Warren, Hewett \& Osmer 1991; Schneider, Schmidt \& Gunn 1987, 1989 \& 1991;
McMahon {\etal} 1994; Kennefick {\etal} 1995). Because these
objects were not selected on the basis of absorption properties,
their spectra are included in constructing the selection function,
$g(z)$, which describes the redshift sensitivity of the survey.
In cases where
damped \lya lines are detected the spectra are used for fitting
Voigt profiles as in the previous section. 
We show the spectra
for all 20 QSOs in figure~\ref{f_speclris}, since even intermediate quality data is lacking
for most of these objects. The $1\sigma$ error arrays are overplotted,
though these are difficult to see in most cases because the signal-to-noise
ratio is $\ge25$. In figure~\ref{f_specfit} we show the spectra for the 
confirmed damped systems. 
The damped candidate and confirmed absorbers are discussed individually below. 

\noindent (1) Q0046$-$293, $z_{abs}$ = 3.0589, log {\nhi}=20.1

The redshift was determined from narrow Si II 1526, Fe II 1608, and 
Al II 1670 and the fit is well constrained by the steep sides
of the HI trough.  
This system is not included in the statistical sample as it 
falls below the threshold of log {\nhi}=20.3 \cm2. 

\noindent (2) Q0046$-$293, $z_{abs}$ = 3.841, log {\nhi}=20.1

This system is composed of two components, with the redshifts
determined from C II 1334.  It is also well constrained by the
HI line.   
This system is not included in the statistical sample as it 
falls below the threshold of log {\nhi}=20.3 \cm2.

\noindent (3) Q0057$-$274, $z_{abs}$ = 3.2413, log {\nhi}=19.8

The redshift was determined from C II 1334 and the fit is well
constrained by the sides of the HI trough.  
This system is not included in the statistical sample as it 
falls below the threshold of log {\nhi}=20.3 \cm2.

\noindent (4) PSS J0059$+$0003, $z_{abs}$ = 3.1043, log {\nhi}=20.0

The redshift is well determined from Si II 1526. Though
there is strong \lya absorption
in the blue wing of the damped profile the resulting
log {\nhi} is accurate because the red wing 
places a strong constraint on the fit.

\noindent (5)   Q0201$+$1120, $z_{abs}$ = 3.3848, log {\nhi}=21.3

The damped system in this QSO was studied by White, Kinney \& Becker (1993)
in a spectrum with $\approx$ 10\AA\ resolution.  
We have wavelength coverage
from 5000$-$8000\AA\ which includes several low ionization metal
lines from the damped absorber. Our column density and redshift
measurements are in excellent agreement with those previously published. 

{\noindent
\begin{tabular}{ll}
(6) PC0953$+$4749,& $z_{abs}$=3.403, log {\nhi}$\le$20.9 \\
                  & $z_{abs}$=3.890, log {\nhi}$\le$21.1 \\
\end{tabular}
}

We originally observed this object because of the obvious large absorption
features visible in the published spectrum.  We did not plan to include
it in the statistical sample because (a) the QSO was approximately a magnitude
fainter than the objects we were originally following up and (b) we observed
it because we knew it might have high column damped features.  
When we completed our new survey, we had observed and included
in our statistical sample all of the QSOs with V $\le$ 19.5 and 
z $\ge$ 3.5 that were available at the time.  This object has a 
published V magnitude of 19.5, so we now include it in our statistical sample. 
It is problematic as it is one of the faintest objects 
we observed and we only have wavelength coverage at one
blue setting. The existing QSO spectra have 25\AA\ resolution
which is too low to provide any useful metal line information.
Even though the 1.5\AA\ resolution LRIS spectrum
has sufficient signal-to-noise to fit the HI we are 
unable to accurately anchor the redshifts for these absorbers so 
the column densities quoted are upper limits. 

{\noindent
\begin{tabular}{ll}
(7) PSS J1435$+$3057,& $z_{abs}$=3.26, log {\nhi}=20.0 \\
                   & $z_{abs}$=3.51, log {\nhi}=20.0 \\
                   & $z_{abs}$=3.71, log {\nhi}=20.0 \\
\end{tabular}
}

Previous experience shows that these candidates
are highly unlikely to be
damped. They are not included in the statistical sample
in the analysis that follows.

\noindent (8) PSS J1443$+$2724, $z_{abs}$=4.216, log {\nhi}=20.8

Though taken with the 300 g/mm grating,
the signal-to-noise ratio in this spectrum is high enough to 
determine that this absorber is highly likely to be damped.  
The redshift was determined from the strong, narrow C II 1334 line.

\noindent (9) PC2047$+$0123, $z_{abs}$ = 2.7299, log {\nhi}=20.4

Although the only useful low-ion metal lines, C II 1334 and
Fe II 1608,  were in
the \lya forest portion of the LRIS spectrum, we 
believe the redshift is well determined, since neither
metal line was severely blended.

\noindent (10) Q2050$+$359, NOT DAMPED

Though the feature at $\approx$ 4843\AA\ appears to be damped
at first glance, an examination of the associated metal lines
reveals that it is not. There is a complex
of low and high ionization metals lines that cover the 
redshift range $2.98 \le z \le 3.17$, and 
there is no suitable fit for a damped \lya 
feature with log {\nhi}$\ge$20.0. It is a complex of lower column
density lines. 

\subsubsection{New Lick/AAT Survey for Damped \lya Systems }

Ten additional QSOs with redshifts $z\ge4$ were observed at the Lick 
3-m telescope with the KAST spectrograph or the Anglo-Australian 
Telescope with the RGO spectrograph, at $\approx$ 6\AA\ resolution.
The survey data from these objects are summarized in table 4 and 
the spectra are shown in figure~\ref{f_speclick}.  
PSS J0244$-$0108 is excluded from further analysis because it exhibits 
broad absorption lines
believed to be intrinsic to the quasar.
Four damped candidates were detected in these spectra and their
estimated column densities were determined from the measured 
equivalent widths of the lines. This method is 
discussed in detail in Lanzetta {\etal}~1991 and its application 
to $z > 4$ QSOs is discussed in SMIH96.
BRI B0111$-$2819 was subsequently followed up with LRIS and is included 
in table~\ref{t2}. 

\section{Statistical Sample}

The statistical sample of damped absorbers includes all confirmed or
candidate damped \lya absorbers with neutral hydrogen column densities 
{\nhi} $\ge 2 \times 10^{20}$ atoms cm$^{-2}$ from surveys designed
to detect these absorbers (Wolfe {\etal} 1986; Lanzetta {\etal} 1991;
LWT95; WLFC95; SMIH96; Storrie-Lombardi, Irwin \& McMahon [SIM96]; 
Storrie-Lombardi \& Hook 2000; this paper). The QSOs with damped absorbers 
are listed in table~\ref{t5}. 
The QSO spectra that did not contain damped absorbers, and the redshift
path they contribute, are listed in table~\ref{t6}.
It includes 646 QSO lines of sight and 85 damped \lya absorbers.
The spectra for all of these lines of sight had sufficient signal-to-noise
to detect damped absorption with {\nhi} above the threshold at
the 5$\sigma$ confidence level between the redshifts $z_{min}$ and 
$z_{max}$ listed in the tables.
The redshift path surveyed in each QSO is used to construct the
sensitivity function, $g(z)$. It gives the number of lines of 
sight at
a given redshift over which damped systems can be detected 
at a $> 5\sigma$ level.
The form of the selection function is given by
\begin{equation}
g(z) = {\sum_{i=1}^{m}}{\Theta}(z_{i}^{max} - z){\Theta}(z - z_{i}^{min})
\end{equation}
where ${\Theta}$ is
the Heaviside step function and where the sum is over the
$m$ QSOs in the survey (see Lanzetta {\etal} 1991).

The upper panel of figure~\ref{f_gz} illustrates how the redshift 
sensitivity 
for the entire statistical sample compares with previous work. 
The data included in this paper are shown as a solid line, 
the large compilation of damped systems in the Large Bright Quasar Survey 
(WLFC95) which provides the bulk of the data for $z < 3$ is shown as a dashed line, 
and the APM survey (SMIH96)
which provided most of the previous data for 
$z > 3$ is shown as a dotted line. 
The lower panel of figure~\ref{f_gz} shows the detail 
for the redshift range $z \ge 3$.

The new data included in this paper increase by 75 percent 
the redshift path surveyed for $z \ge 3$.
In table~\ref{t7} we compare $\Delta z$ for the LBQS survey, the APM
survey, and this paper for $z>3$ and for the entire redshift
range of the samples. 
We also give analogous values for the total ``absorption distance''
$\Delta X$ which is defined as 
\begin{equation}
X(z) = \cases {{2 \over 3}[(1+z)^{3/2} - 1] &if q$_0=0.5$;\cr
              {1 \over 2}[(1+z)^2 - 1]    &if q$_0=0$.}
\label{absdis2eqn}
\end{equation} (\cite{BP69}; cf.~\cite{Tytler87}).
The redshift path over which damped systems could be detected is 
crucial in estimating the cosmological mass density in neutral gas 
from the damped systems, as well as the number density per 
unit redshift, and the frequency distribution per unit column density.
These are discussed in detail in the next section.

\section{Analysis}

\subsection{Frequency Distribution of Column Densities, $f(N,z)$}
 
By combining our statistical sample of damped systems 
with the data presented in Hu {\etal} (1995) for column densities
{\nhi} $\le 2 \times 10^{20}$ atoms cm$^{-2}$,  
we can examine the frequency distribution for the entire 
HI column density range, $10^{12} - 10^{22}$ atoms cm$^{-2}$
(see \cite{Tytler87}; \cite{Petitjean93}; \cite{Hu95}).
This is shown in figure~\ref{f_fnallhi}. To first order 
it is fit by a power law function,
$f(N) \propto N^{-1.46}$, over ten orders of magnitude
in column density. The \lya forest absorbers are by
far the most numerous, yet if we integrate this to
find the mass density, the bulk of the HI mass is
locked up in the damped systems.
A disadvantage of a power law model for the HI column density 
distribution of damped Ly$\alpha$
absorbers is the divergent nature of the integral mass contained in the
systems. A high column density cutoff (\eg\ $10^{22}$) must be selected
to do the integral. We discuss below  
an alternative parameterization based on a gamma distribution
(\cf\ Pei \& Fall 1995; SIM96).

\subsubsection{Differential Column Density Distribution}
\label{s_diffdx}

The frequency distribution of HI column densities is defined as
follows.  Let $f(N,X){\rm d}N{\rm dX}$ be the number of absorbers
per sightline with HI column densities in the interval $(N,N+{\rm d}N)$,
and absorption distances in the interval $(X,X+{\rm d}X)$.
Then the frequency distribution of HI column densities
$f(N,z) \equiv f(N,X[z])$. 
In figure~\ref{f_fnalldiff} we show the $f(N,z)$ derived from 
the entire statistical sample
of damped absorbers for $q_{0}$ = 0.0, and in 
figure~\ref{f_fnsplitdiff} we show the 
same data divided into four redshift bins: 
$z = 0.008-1.5$, $z = 1.5-2.5$, $z = 2.5-3.5$, 
and $z = 3.5-4.7$. The flattest distribution is 
in the redshift range $z = 3.0-3.5$, and it steepens towards higher
and lower redshifts from this point.
This was noted in WLFC95 for the $z < 3.5$ data and in SIM96 
for the $z > 3.5$ data, but the differences in f(N) with redshift
were not statistically significant.  Our new statistical sample
is large enough to make some quantitative statements about the redshift
evolution of f(N) at high redshift.

\subsubsection{The Neutral Gas Density at z $<$ 1.5}

There is a steepening of the differential distribution of 
column densities evident at z $<$ 1.5 in our data set but 
not at a statistically significant level.
Previous surveys have shown a very lower number density 
per unit redshift from International Ultraviolet Explorer and
Hubble Space Telescope (HST) data (LWT95) and 
Jannuzi \etal\ (1998), also find numbers consistent with this 
from HST Key Project data. But Rao \& Turnshek (1998) 
have reported in a Letter the discovery of 12 new low 
redshift damped systems and their full survey has just
appeared in preprint form (Rao \& Turnshek 2000). 
This redshift range is problematic because any bias in QSO surveys 
due to possible obscuration by dust in foreground absorbers
would have the largest impact here (see Pei \& Fall 1995; SMI96).
We focus in this paper on the redshift range z $>$ 1.5.

\subsubsection{Evolution of the Neutral Gas Density at z $>$ 1.5}
\label{s_steeper}

For the first time we determine a {\it statistically significant}
($>$ 99.7\% confidence) steepening in the column density distribution
function at redshifts z $>$ 4.0.  This is evidenced by 
comparing the data in the redshift range 
1.5 $<$ z $<$ 4 with the data at z $\ge$ 4 as shown
in figure~\ref{f_gamcumks}. This shows the cumulative distribution,
normalized by the absorption distance surveyed.
A Kolmogorov-Smirnov (K-S) test gives a probability of only 0.006
that the two redshift samples are drawn from the same distribution.
The steepening of the distribution function is
due to both fewer very high column density absorbers
(\nhi $\ge$ $10^{21}$ \atomcm) and more lower column density
systems (\nhi $= 2-4 \times 10^{20}$).
No damped systems with column densities log \nhi $\ge$ 21 have 
yet been detected at $z > 4$. 

\subsubsection{Fits to the Column Density Distribution }
 
A single power law, f(N) $=$ kN$^{-\beta}$
does not provide a good fit to the column density distribution
function for damped \lya absorbers. A single power law
fit has the additional problem that if $\beta < 2$,
as all current estimates indicate, then the total mass in damped systems
diverges unless an upper bound to the HI column density is assumed.
An alternative parameterization using a gamma
function to describe the HI column density distribution 
was suggested by Pei \& Fall (1995) and adopted by
SIM96. We use the same formalism here.  

We model the data with a gamma distribution of the form
\begin{equation}
f(N,z)=( f_* / N_* ) ( N / N_* )^{-\beta} e^{-N/N_*}
\label{gamdisteqn}
\end{equation}
where $f_*$ is the characteristic number of absorbing systems at the column
density $N_*$, and $N_*$ is a parameter defining the turnover, or `knee',
in the number distribution.  Both $f_*$ and $N_*$ may in general vary with
redshift but for the moment we treat them as constants.  This functional
form is similar to the Schechter luminosity function (Schechter 1976).
For $N << N_*$ the gamma function tends to the same form as the single power
law, $f(N) \propto N^{-\beta}$; while for $N$ \simgt\ $N_*$, 
the exponential term begins to dominate.
We use a maximum likelihood technique to find a solution
over a two-dimensional grid of pairs of values of $N_*$ and
$\beta$, since the constant $f_*$ can be directly computed
using the constraint
\begin{equation}
m =\sum_{i=1}^n f_* \int_{N_{min}}^{N_{max}} \int_{z_{min}^i}^{z_{max}^i} f(N,z)
 dz dN
\label{gamconstrainteqn}
\end{equation}
where m is the total number of observed systems.

The results of fits to the data in the redshift ranges
1.5 $<$ z $<$ 4 and z $\ge$ 4, overplotted on the cumulative
distribution of absorbers, are shown in the upper panels 
of figure~\ref{f_gamcum}. 
A data point for the expected number of Lyman-limit systems that
would be detected down to $\log$ \nhi $=17.2$ is shown with a circled star.
Lyman-limit systems are defined and
detected by the observation of neutral hydrogen (HI) absorption
which is optically thick to
Lyman continuum radiation for $\lambda <$ 912\AA, the Lyman limit,
corresponding to a column density {\it N}(HI)$\ge 1.6 \times 10^{17}$ cm$^{-2}$
(See Tytler 1982; Sargent, Steidel \& Boksenberg 1989; Lanzetta 1991; 
Storrie-Lombardi \etal\ 1994; Stengler-Larrea \etal\ 1995 for discussions of 
Lyman Limit systems.)
Their contribution is calculated by integrating the number density per unit
redshift of Lyman limit systems expected ($N(z)=0.27(1+z)^{1.55}$) over the 
redshift path covered by the $n$ QSOs in the damped sample, \ie\
\begin{eqnarray*}
{\rm LLS}_{expected} &=& \sum_{i=1}^n \int N(z)dz =
         \sum_{i=1}^n \int_{z_{min}}^{z_{em}} N_0(1+z)^{\gamma}dz \\
  &=&  \sum_{i=1}^n \int_{z_{min}}^{z_{em}} 0.27(1+z)^{1.55}dz
\end{eqnarray*}
\begin{equation}
\label{expllseqn}
\end{equation}
(SMIH94,SIM96).
Including the Lyman limit system point provides a longer baseline 
in column density and allows us to make better estimates of the 
uncertainties in the fit. 
The log-likelihood function results with confidence
contours are shown in the lower panels of figure~\ref{f_gamcum}.

The parameters for the fit in equation~\ref{gamdisteqn} 
are very similar for both redshift ranges when the Lyman limit
systems are included, with log N$_* = 21.22$ \atomcm\ 
for both and $\beta = 1.24$, f$_* = 0.05$ for the lower redshift
data and $\beta = 1.08$, f$_* = 0.16$ for the higher redshift
data. 
Without the Lyman-limit point anchoring the lower column density 
end, the fits to just the high column density damped systems yield 
$\beta = 1.52\pm0.25$, f$_* = 0.03$, and 
log N$_* = 21.39\pm0.30$ for the lower redshift data and
$\beta = 0.32\pm0.6$, f$_* = 0.34$, and log N$_* = 20.92\pm0.50$ 
for the higher redshift data.
The errors on the exponent $\beta$ are very large for the 
z $>$ 4 data due to the small number of data points.
The fits reflect what we saw in looking at the differential 
counts, \ie\ the ``knee'', N$_*$, where the distribution turns over
is at a lower column density for the higher redshift data. 
The best fit results are tabulated in table 8.

In SIM96, with a much smaller data set (366 QSOs, 44 absorbers),
we obtained log $N_*=21.63\pm0.35$, $\beta=1.48\pm0.30$, 
and $f_*=1.77\times 10^{-2}$ fitting the whole sample.
With our current statistical sample (646 QSOs, 85 absorbers)
we find very similar results with  
$N_*=21.33\pm0.30$, $\beta=1.48\pm0.25$, 
and $f_*=3.19 \times 10^{-2}$.
Alternatively, single power law fits of the 
form f(N) $\propto$ N$^{-\beta}$ (not including the point for the 
Lyman-limit systems)  
yield $\beta = 1.96$ for z $>$ 4 and $\beta = 1.78$ for 1.5 $<$ z $<$ 4,
again consistent with values obtained in earlier work (LWTLMH91; SIM96).

\subsection{Number Density Evolution with Redshift}

We now examine the redshift
evolution of the damped \lya absorbers in our combined data set
by determining the  number density of absorbers
per unit redshift, $dN/dz \equiv N(z)$.  
In a standard Friedmann Universe for
absorbers with
cross-sectional area A, absorption distance $X$, and
$\Phi_{0}$ per unit comoving volume,
\begin{equation}
%N(z) = \Phi_{0}\pi{R_{0}}^2cH_{0}^{-1}(1 + z)(1 + 2q_{0}z)^{-1/2}.
N(z) = \Phi_{0}AcH_{0}dX/dz
\label{sybteqn}
\end{equation}
where
\begin{equation}
dX/dz = \Big[ {{(1+z)^2} \over (1+z)^2 (1+\Omega_M z) - z(z+2) \Omega_{\Lambda}}\Big]^{1/2}.
\label{sybteqn2}
\end{equation}
It is customary to represent the number density as a power law of the form
\begin{equation}
N(z) = N_{0}(1 + z)^\gamma,
\label{nzeqn}
\end{equation}
The differential distribution in number density of absorbers versus
redshift is shown in figure~\ref{f_dndz_all}.  It is fit by a
single power law with $N_{0} = 0.055$ and $\gamma = 1.11$, which
would suggest no intrinsic evolution in the product of the
space density and cross-section of the absorbers with redshift,
but the fit is very poorly constrained. This is evident in the
$>$68.3\% and $>$95.5\% confidence contours for
the log-likelihood results when calculating $\gamma$ and $N_0$
which is shown in figure~\ref{f_maxz}.  The poor fit is due to differential
evolution in the number density of damped \lya absorbers with different
column densities which has been described in LWT95, WLFC95, and SIM96.
 
The number density of absorbers versus redshift, split
at a column density of log {\nhi} = 21 is shown in
figure~\ref{f_dndz_split}.
The number density of systems with log {\nhi} $>$ 21,
shown as solid lines, peaks at z $\approx$ 3.5,
when the Universe is 15-20\% of its present age. These
systems then disappear at a much faster rate from
z$=$3.5 to z$=$0 than does the population of damped absorbers as a whole.
There is a paucity of very high column density systems at
the highest redshifts surveys, which was also evidenced in the steepening
of the column density distribution discussed in $\S$~\ref{s_steeper}.
The number density of damped absorbers with column densities
log {\nhi} $\le$ 21 decreases from redshifts z $\approx$ 4 to z $\approx$ 3.5
and remains relatively constant towards z = 1.5.
The differential evolution with column density suggests:
\begin{enumerate}
\item There has been insufficient time at z $>$ 3.5 for the highest column
density absorbers to collapse.
\item Once they do form, the highest column density absorbers
preferentially form stars before their lower column density counterparts,
and hence disappear more rapidly towards lower redshifts. 
\end{enumerate}

From the evolution of the HI with redshift alone
we are unable to determine if we are watching the evolution
of similar systems with redshift or
watching some systems disappear and others form.
 
\subsection{Evolution of the Mass Density of Neutral Gas} 

The comoving mass density of neutral gas
is given by
\begin{equation}
\Omega_g(z) = {\frac{H_{0}}{{c}}}{\frac{{\mu}m_{H}}{\rho_{crit}}}{\frac{\sum_{i}N_{i}({HI})}{\Delta X(z)}}
\end{equation}
where the density is in units of current critical
density (see Lanzetta \etal\ 1991). The quantity $\mu$ is 
the mean particle mass
per $m_{H}$ where the latter is the mass of
the H atom, $\rho_{crit}$ is the current critical mass density, 
and the sum is over damped \lya systems in the statistical
sample listed in table~\ref{t5}.  We use H$_0$ = 65 {\kmsM}
in all these calculations.
Because we are interested in determining $\Omega_{g}(z)$ for
separate redshift bins, $\Delta X(z)$
is given by the integral $\int g(X)dX$ between
the $X(z)$ corresponding to the redshift limits of each bin. We determine
$\Omega_{g}(z)$ by assuming $\mu$ = 1.3.
The results for the our data set from z $=$ 0.008 to z $=$ 4.7
are shown in figure~\ref{f_omega} for 
3 different cosmologies with $\Omega$ = \Wmatter + \Wlamb:
\begin{itemize}
\item \Wmatter=1, \Wlamb = 0
\item \Wmatter=0.3, \Wlamb = 0
\item \Wmatter=0.3, \Wlamb = 0.7
\end{itemize}
The region {\Wstar} is the $\pm1\sigma$ range for the
mass density in stars in nearby galaxies
(Fukugita, Hogan \& Peebles 1998).  The point at z$=$0 is
the value inferred from 21 cm emission from local galaxies
%(Fall \& Pei 1993; Rao \& Briggs 1993).  These results
(Zwaan \etal\ 1997).  These results
are tabulated in table~\ref{t9}. 
{\Wg} evolves in the same fashion with redshift regardless of
the cosmology selected and there is a tentative peak
in {\Wg} in the redshift range 3 $<$ z $<$ 3.5.
We consistently find very high column density systems in
this redshift range, even though for z $>$ 4 QSOs, z$\sim$ 3
absorbers are deep in the \lya forest where they are most
difficult to detect due to the density of the forest lines
in the spectra.  This was evidenced in section \S~\ref{s_diffdx} 
where the shape of the frequency distribution of column densities
was flattest in the  3 $<$ z $<$ 3.5 redshift bin.

What does change with cosmological model
is the ratio of the peak value of {\Wg} to \Wstar,
the mass density in galaxies in the local universe.
We define this ratio as {\Rgstar} $\equiv$ \Wg($3<z<3.5)/$\Wstar.
Previous surveys (Lanzetta \etal\ 1991; WLFC95; SMI96)
have found the value of {\Rgstar} to be of order 1.
When we look at this ratio ({\Rgstar} = \Wg($z=3$)$/$\Wstar)
For our current data set and value of {\Wstar},
we find {\Rgstar} $\approx$ 0.25 - 0.5, depending on the cosmology.
For an $\Omega = 1$ Universe with a zero cosmological constant
{\Rgstar} $\approx$ 0.5,
for an $\Omega = 1$ Universe with a positive
cosmological constant (\Wlamb = 0.7, \Wmatter = 0.3)
{\Rgstar} $\approx$ 0.25, and
for a universe with \Wlamb = 0 and \Wmatter = 0.3,
{\Rgstar} $\approx$ 0.3.

\section{Discussion}

Though the results for the comoving mass density of neutral 
gas shown in figure~\ref{f_omega} look 
very similar to those published in previous work we discuss
below the main differences in these new results and 
their implications. 

\begin{enumerate}
\item The values of {\Wg} in this work
are lower than those previously published for $z > 1.5$.
The values for {\Wg} published in WLFC95 and SMI96 used 
H$_0$ = 50 {\kmsM} and in this work we use H$_0$ = 65 {\kmsM}
which decreases {\Wg} by 23\% at all redshifts. 
The improved statistics due to the increased redshift path
also reduces the effect of the scatter produced by the small 
number of QSOs with very high column density absorbers.
The value of {\Wg} is now less sensitive to the maximum column
density in any particular redshift bin. 
The comparison of {\Wg} to {\Wstar} does not appear very different, even though
the values are lower, because our estimate of {\Wstar}
has changed as well.   

\item The value of {\Wstar} from Fukugita, Peebles \& Hogan (1998)
that we are using in these plots is also about half the
value inferred from Gnedin \& Ostriker (1992) used in previous
determinations.  There are still substantial uncertainties
in the relative contributions of disks and spheroids in
determining the local mass density (e.g.~Salucci \& Persic 1999;
J. Dalcanton, R. Marzke and L. Yan, private communication).  Salucci 
\& Persic (1999) find of lower value of {\Wstar} ($2 \times 10^{-3}$)
independent of the Hubble constant, which is about half the value
we have plotted from Fukugita, Peebles \& Hogan (1998). As our
understanding of the local mass density improves we will 
continue to refine these numbers for comparison with the high
redshift determinations. 

\item The calculation of {\Wg} for z $>$ 3.5 in SMI96 hinted
at a turnover at these high redshifts. Our larger data set
does not confirm whether or not this turnover is real though
the trend continues to be present in our larger dataset. 
We do see a statistically significant change in the frequency
distribution of column densities at $z > 4$
(\S~\ref{s_steeper}),
which suggests we are probing an epoch when the highest column
density absorbers have not yet collapsed. Also, for one
of the highest column density high redshift systems
(log {\nhi} $\le$ 21.1 at $z = 3.89$) from PC0953$+$4749,
we only have an upper limit for the column density as
discussed in section (\S~\ref{s_0953}).  Though we have 
surveyed enough quasars at z $>$ 3.5 to not be extremely 
sensitive to the inclusion or exclusion of one absorber, 
it is true that if this high column density 
were substantially lower, the
error bars for the 3 $<$ z $<$ 3.5 and z $>$ 3.5 bins
would not overlap.  We will have to wait for more data
to make a definitive statement about a turnover in
{\Wg} at z $>$ 4.

\item {\Wg} determined in WLFC95 shows a monotonic decrease
from $z=3.5 \rightarrow 0.008$. Our larger data set also shows
a decrease, but the inclusion of the addition survey of
Storrie-Lombardi \& Hook (2000) makes {\Wg} flatter in the
redshift 2 $<$ z $<$ 3 range.
Our new data are consistent with a constant value of
{\Wg} for $2 < z < 4$.  New results from Rao and
Turnshek (2000) for $z < 1.5$ also suggest \Wg (z$<$1.5) may be higher
 than
previously determined.  In figure~\ref{f_omega_RT} we replot 
the first panel of figure~\ref{f_omega},
and show their lower redshift data point with dashed lines.
The error bars are still very large but they find a substantially
higher value for {\Wg} at z $<$ 1.65 than previous surveys.

\end{enumerate}

\acknowledgements

We thank Julia Kennefick, George Djorgovski, Mike Irwin, and 
Richard McMahon for providing QSO coordinates prior to publication 
and the staff at Keck, Lick, and the AAT for their able assistance 
in obtaining these observations. LSL thanks Jason Prochaska, Celine
Peroux, and Max Pettini, and an anonymous referee for helpful 
suggestions.   

%\clearpage

{}

\clearpage
%%% FIGURES 

%% Figure 1 
\plotone{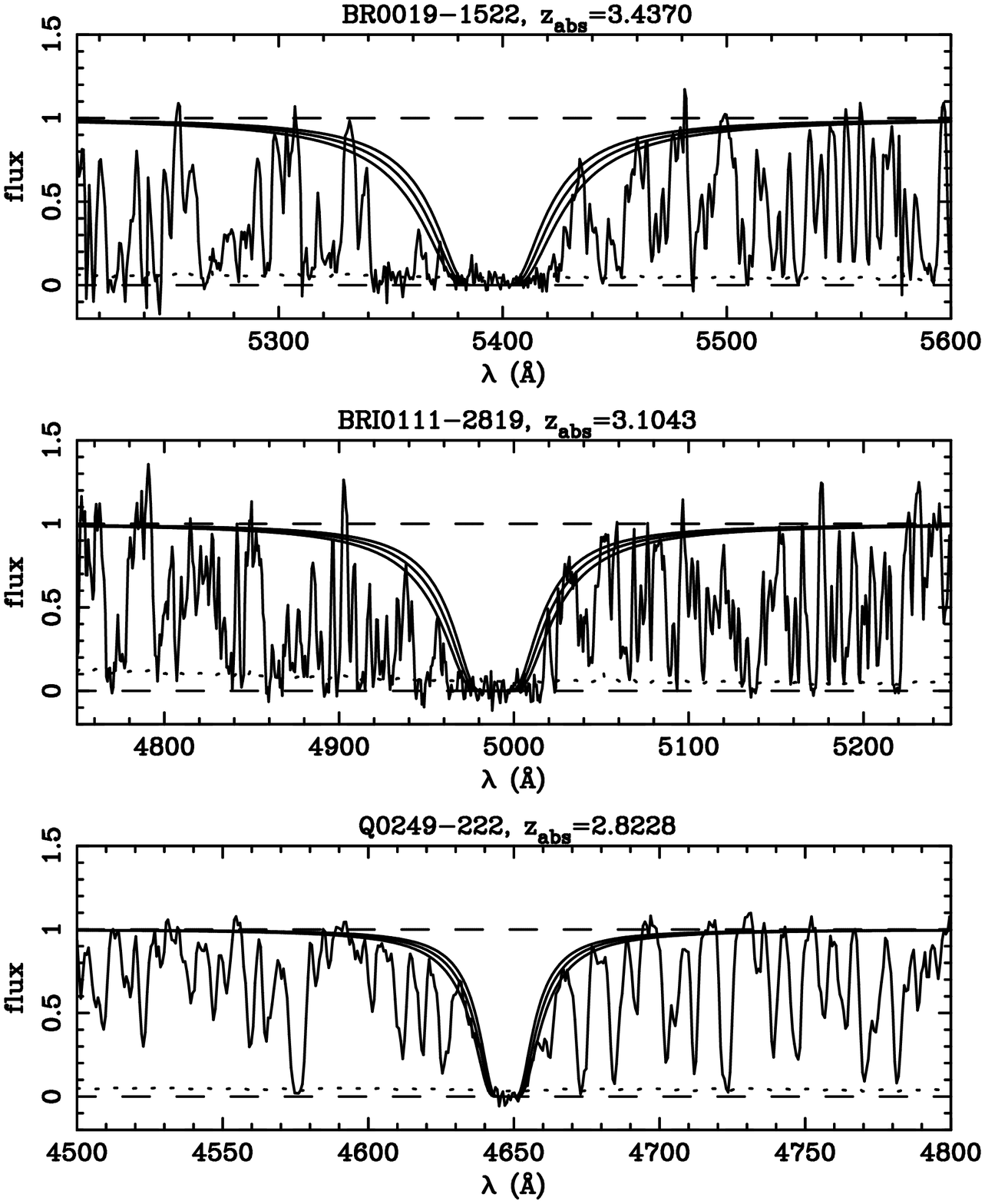}

\plotone{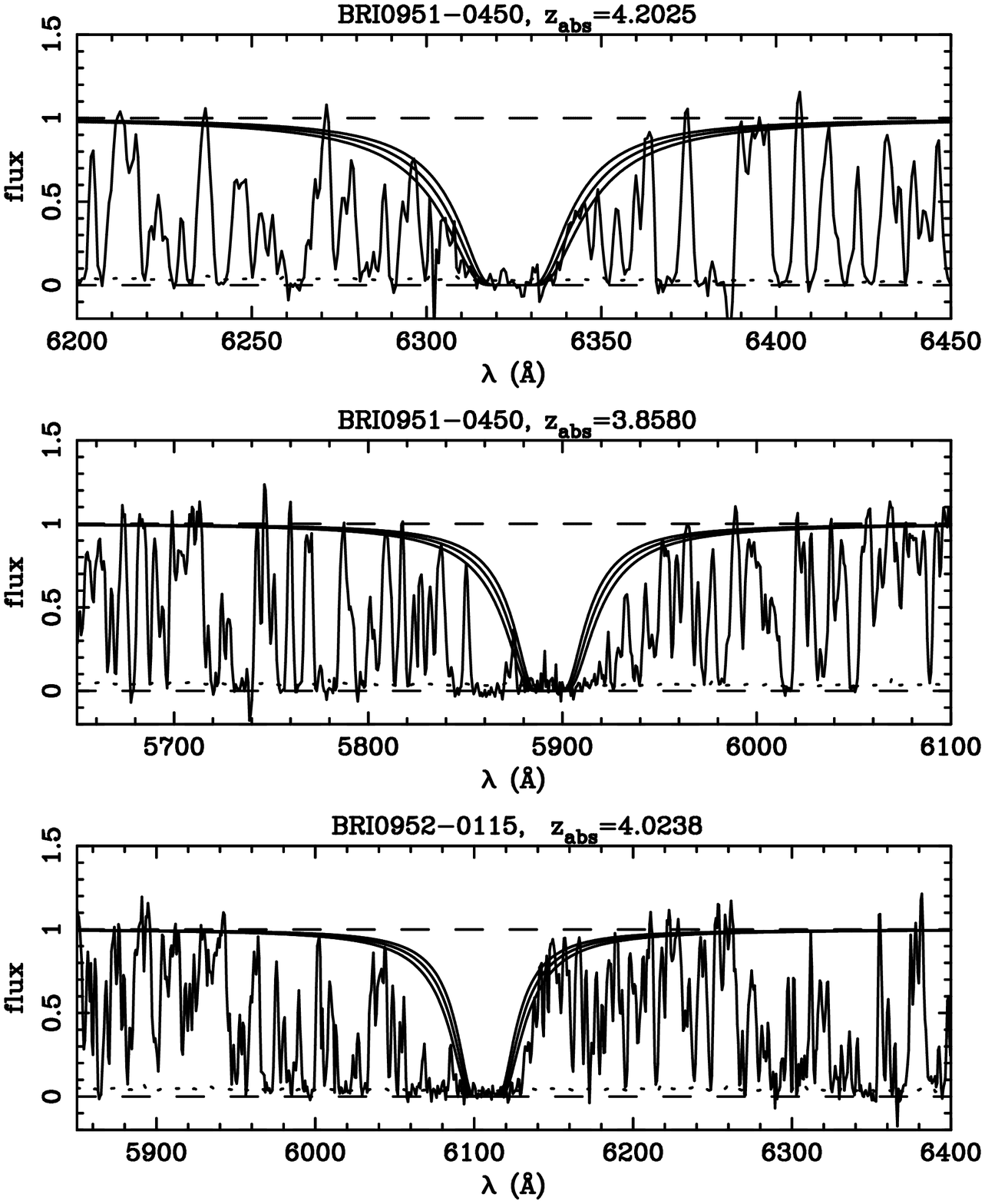}

\plotone{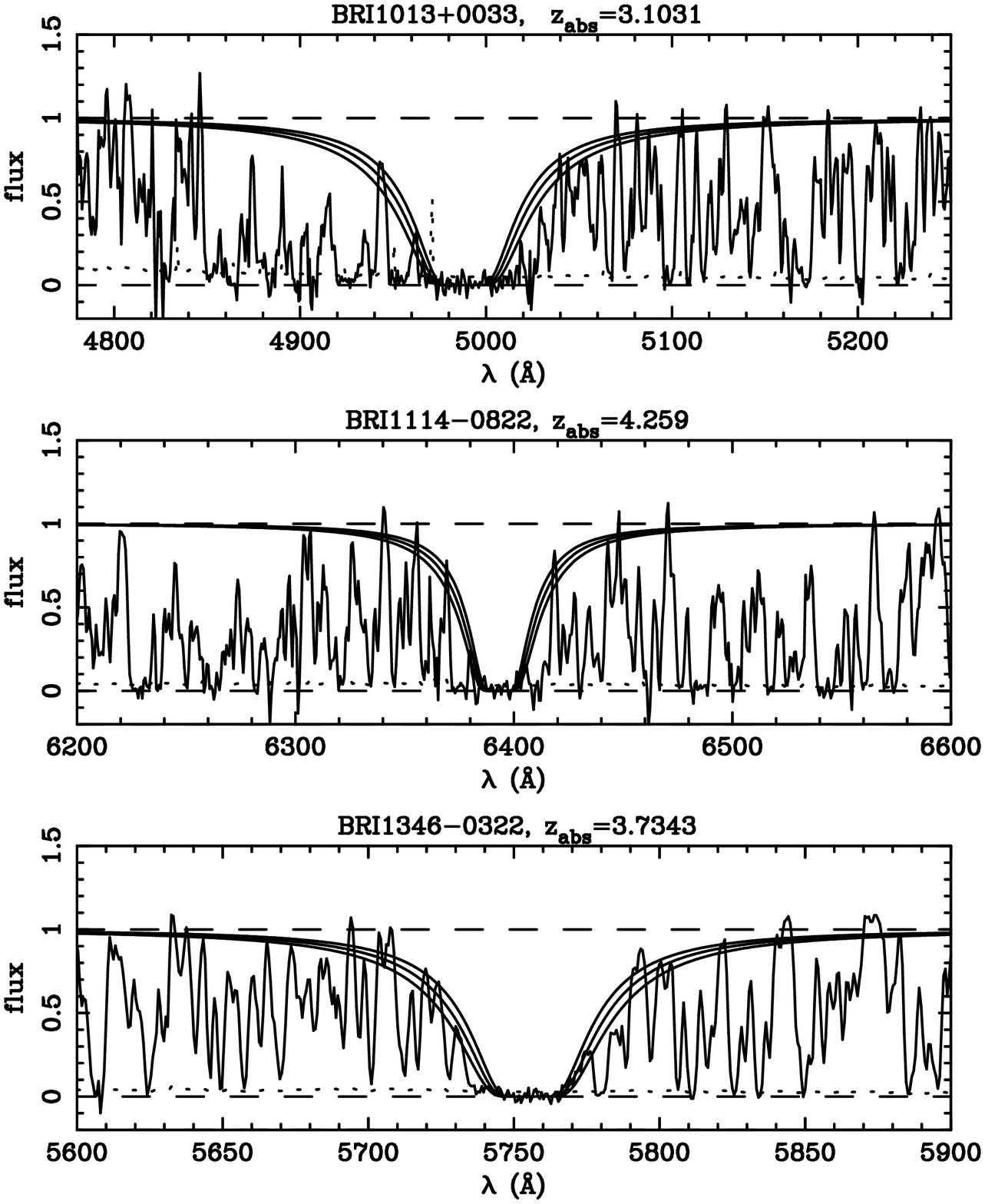}

\plotone{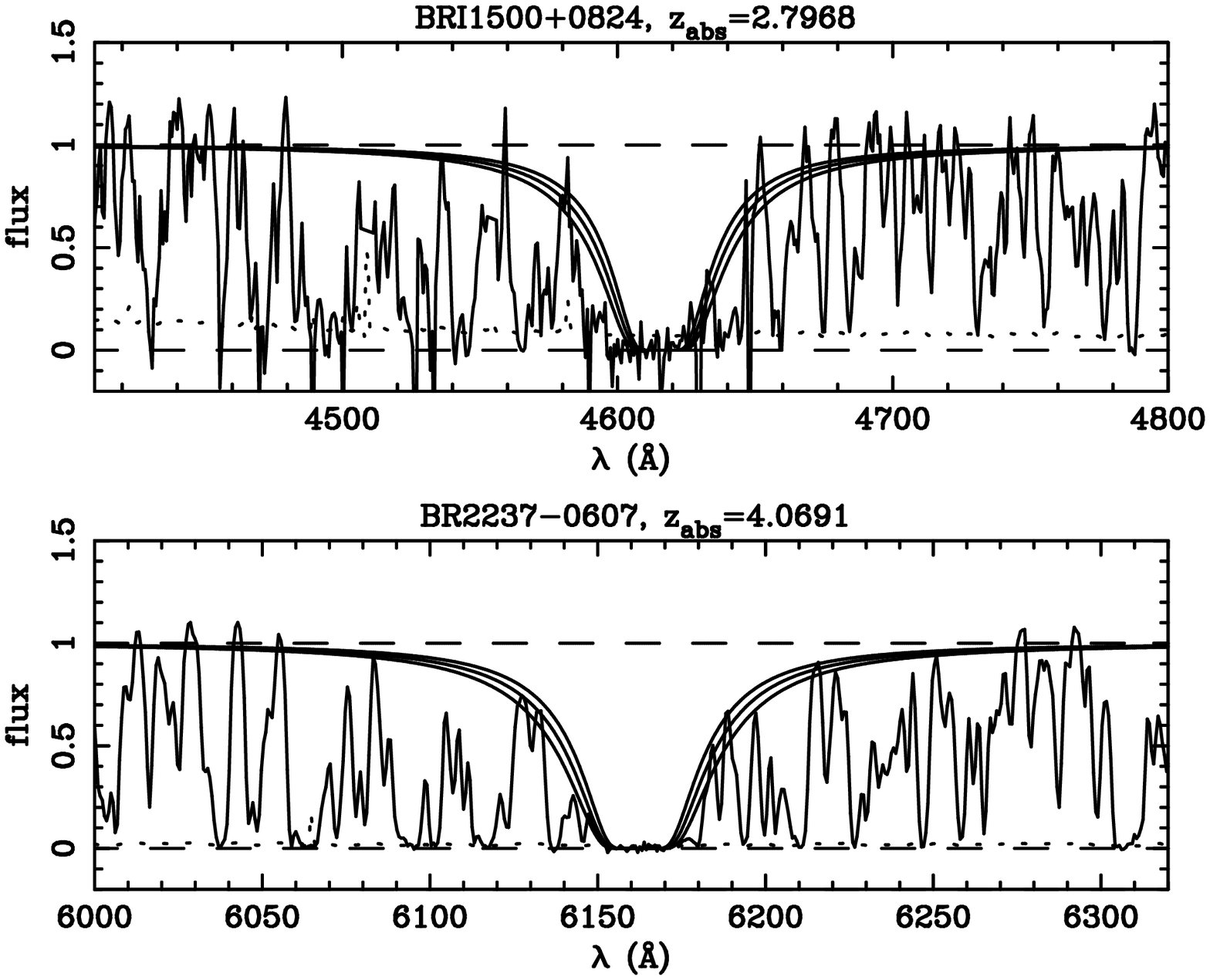}

\figcaption{The spectra for
the 10 confirmed damped \lya systems with log {\nhi} $\ge$ 20.3
cm$^{-2}$ and the $z$ = 2.8228 system toward Q0249$-$2212 in which
log {\nhi} = 20.20 cm$^{-2}$ are plotted.
Three Voigt profile fits are shown
in order of increasing {\nhi} with the middle
value representing the mean {\nhi} inferred from the optimal fit,
and the other two judged to be $\pm 1\sigma$ from the mean. The results
are summarized in table 2.
\label{f_specconf}}

\clearpage
%% Figure 2
\plotone{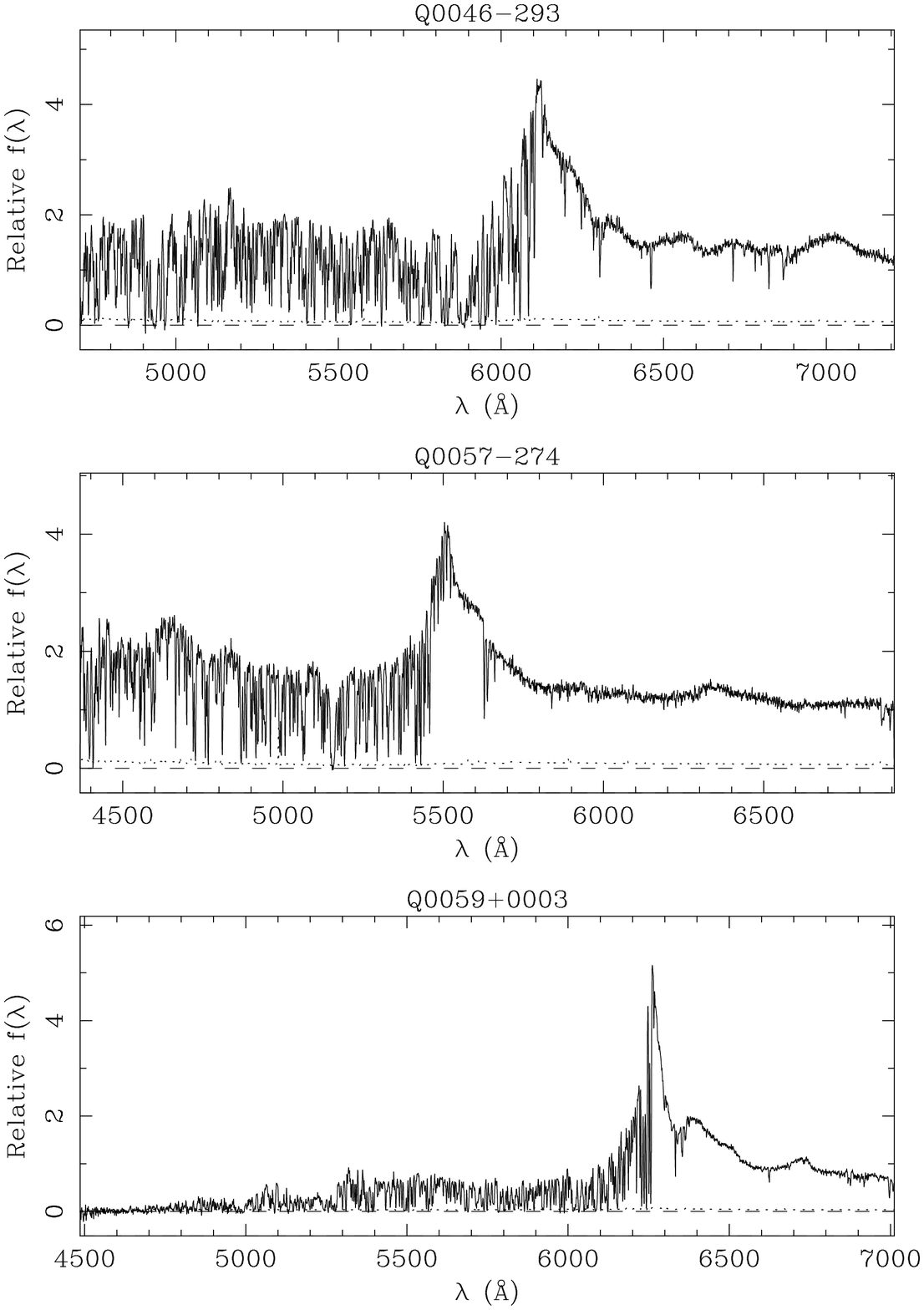}
 
\plotone{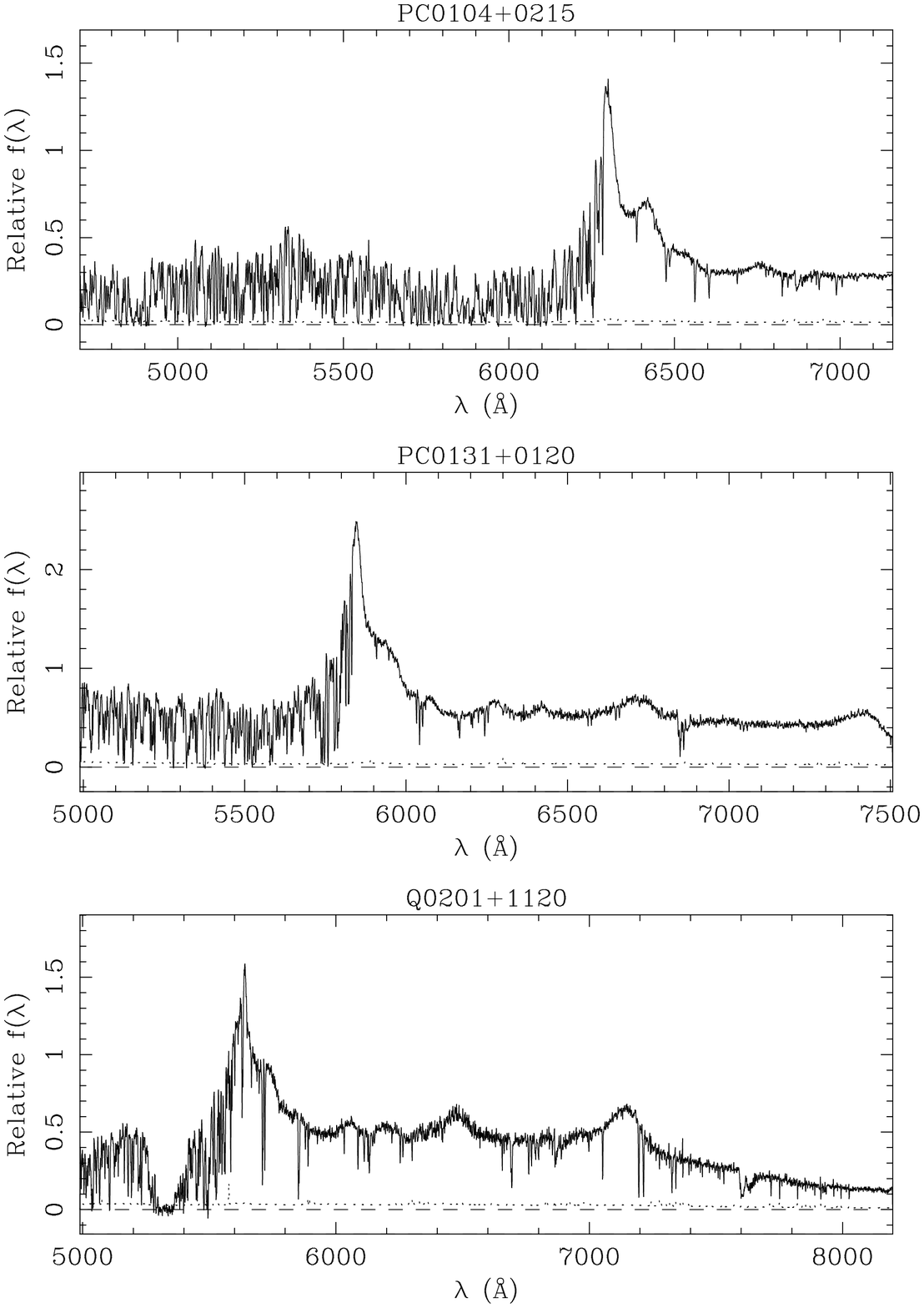}
 
\plotone{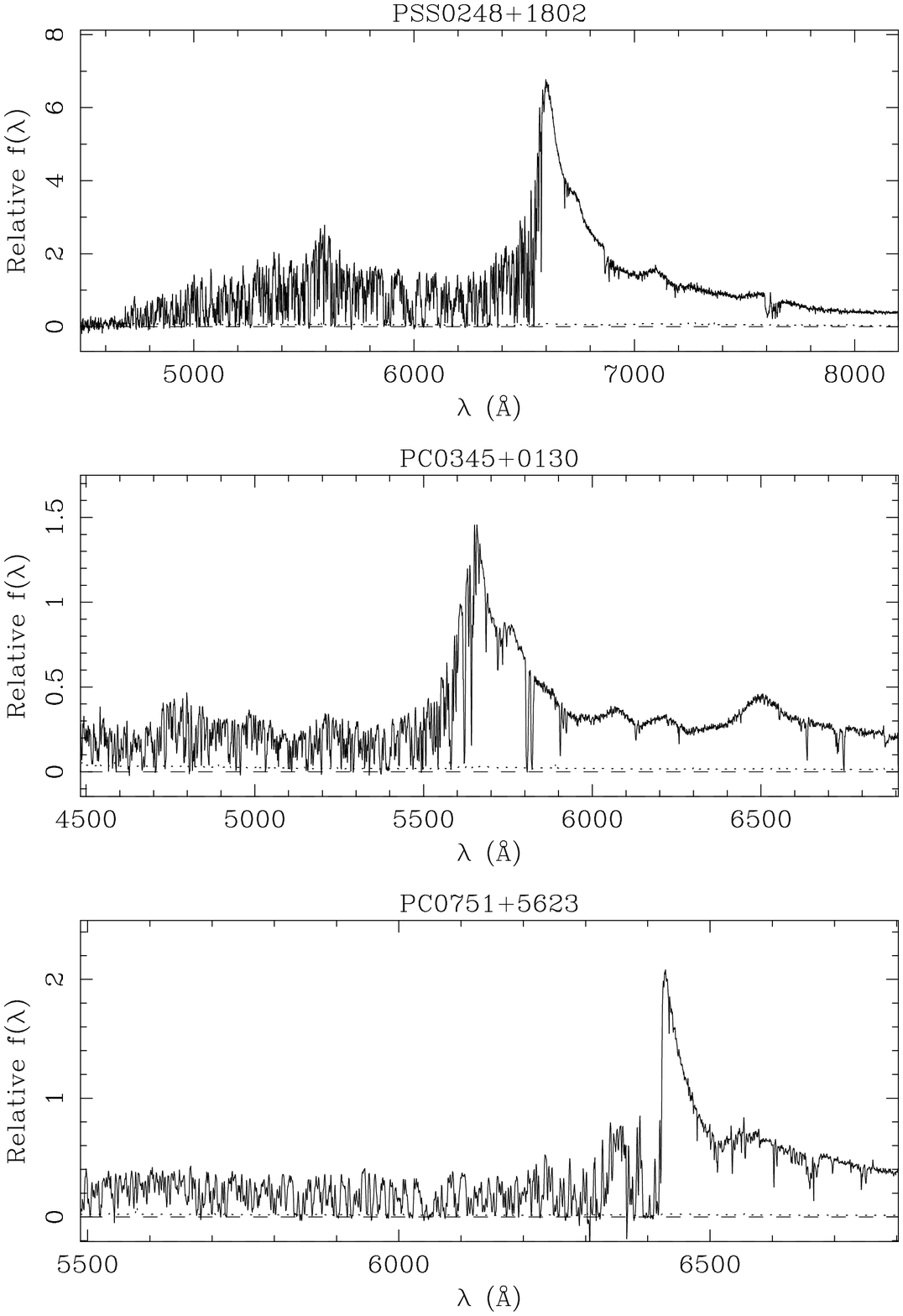}
 
\plotone{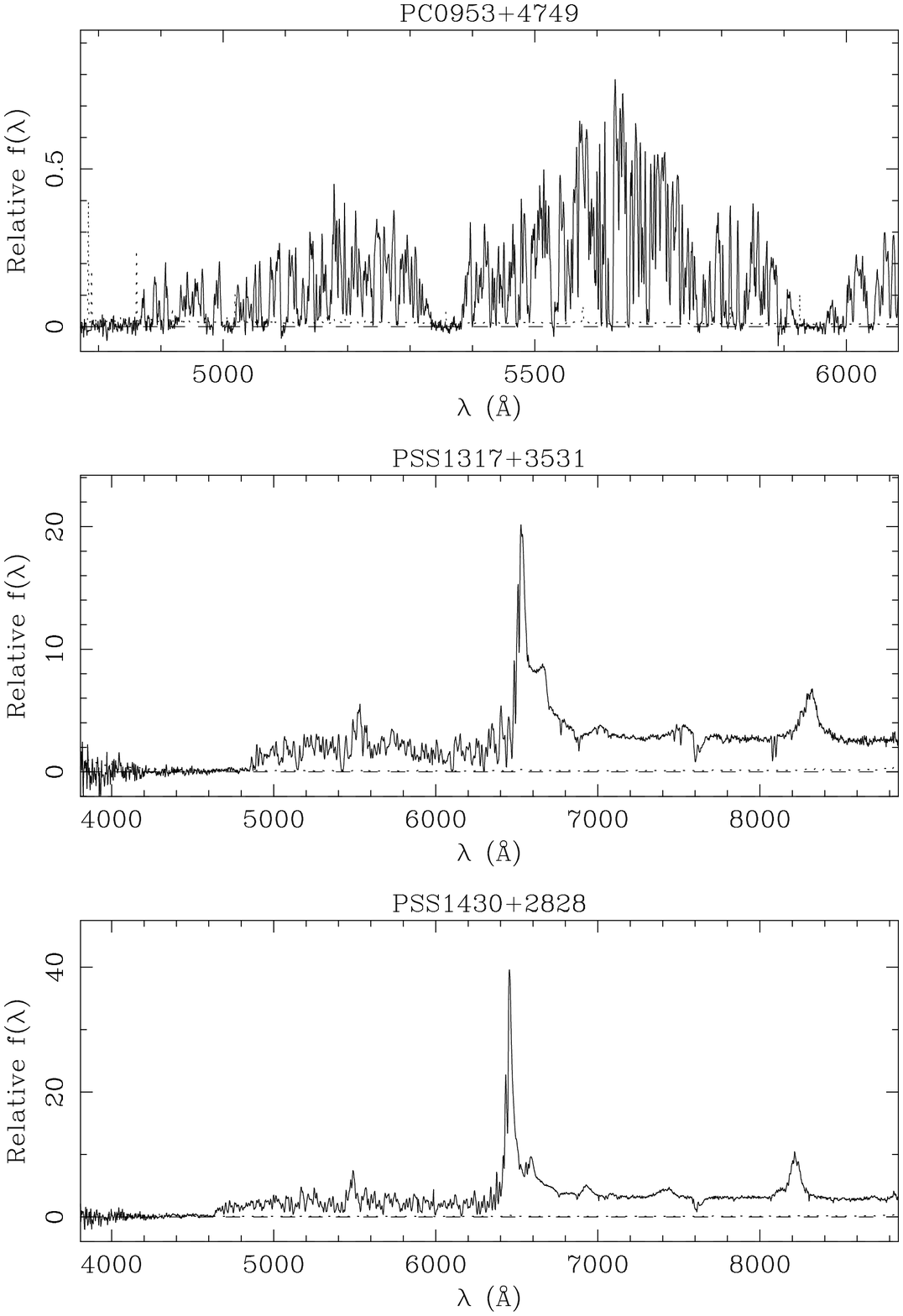}
 
\plotone{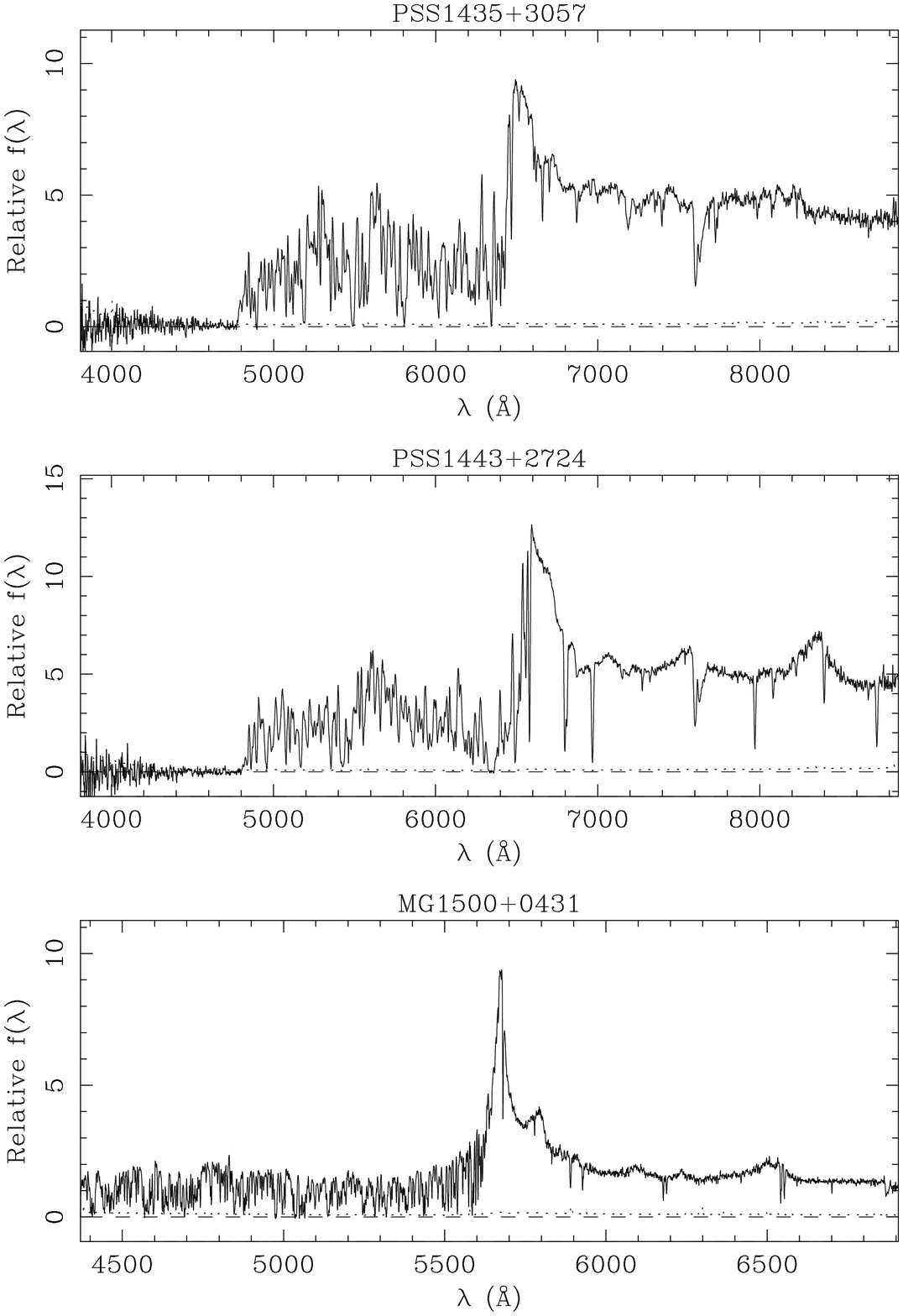}
 
\plotone{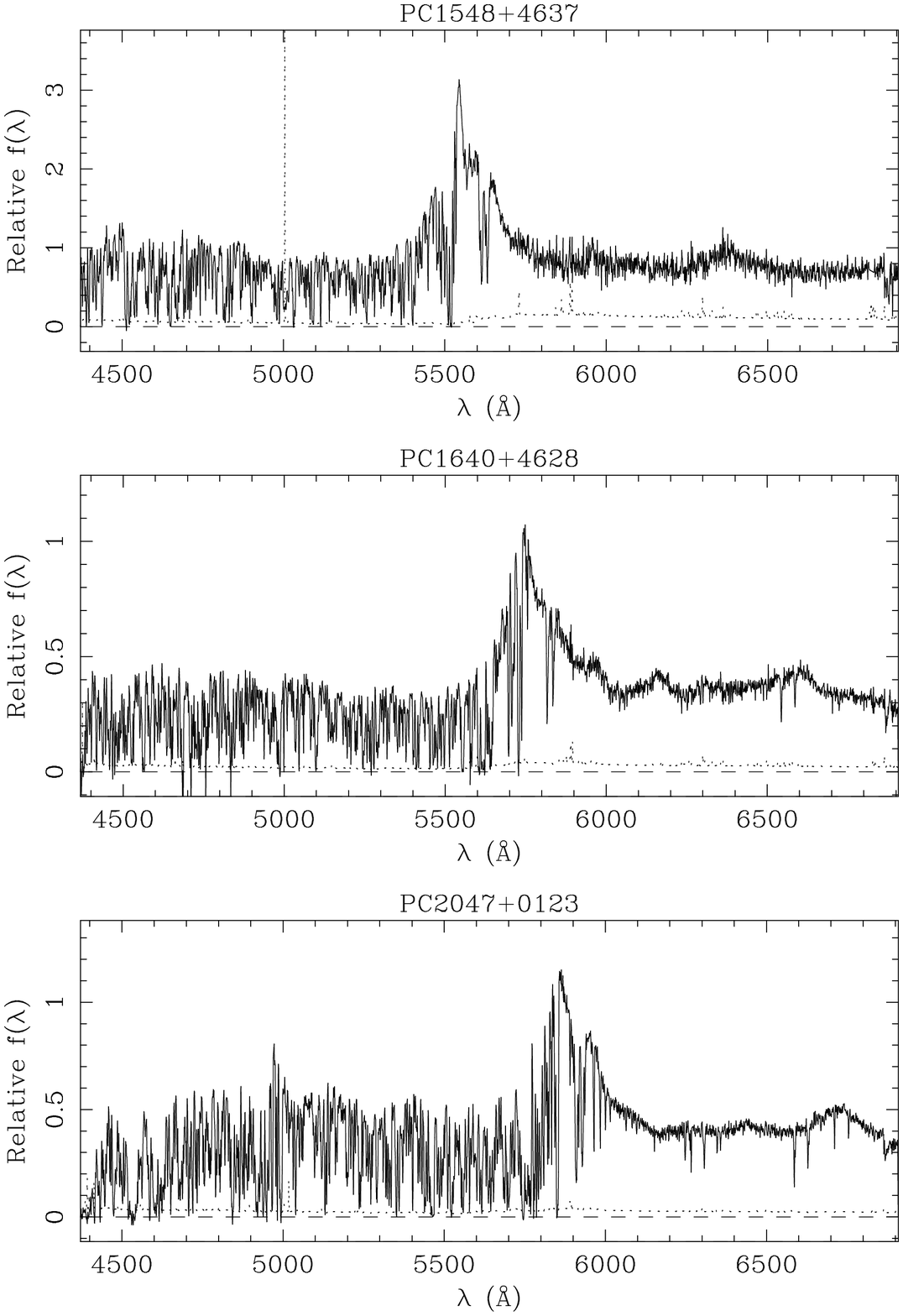}
 
\plotone{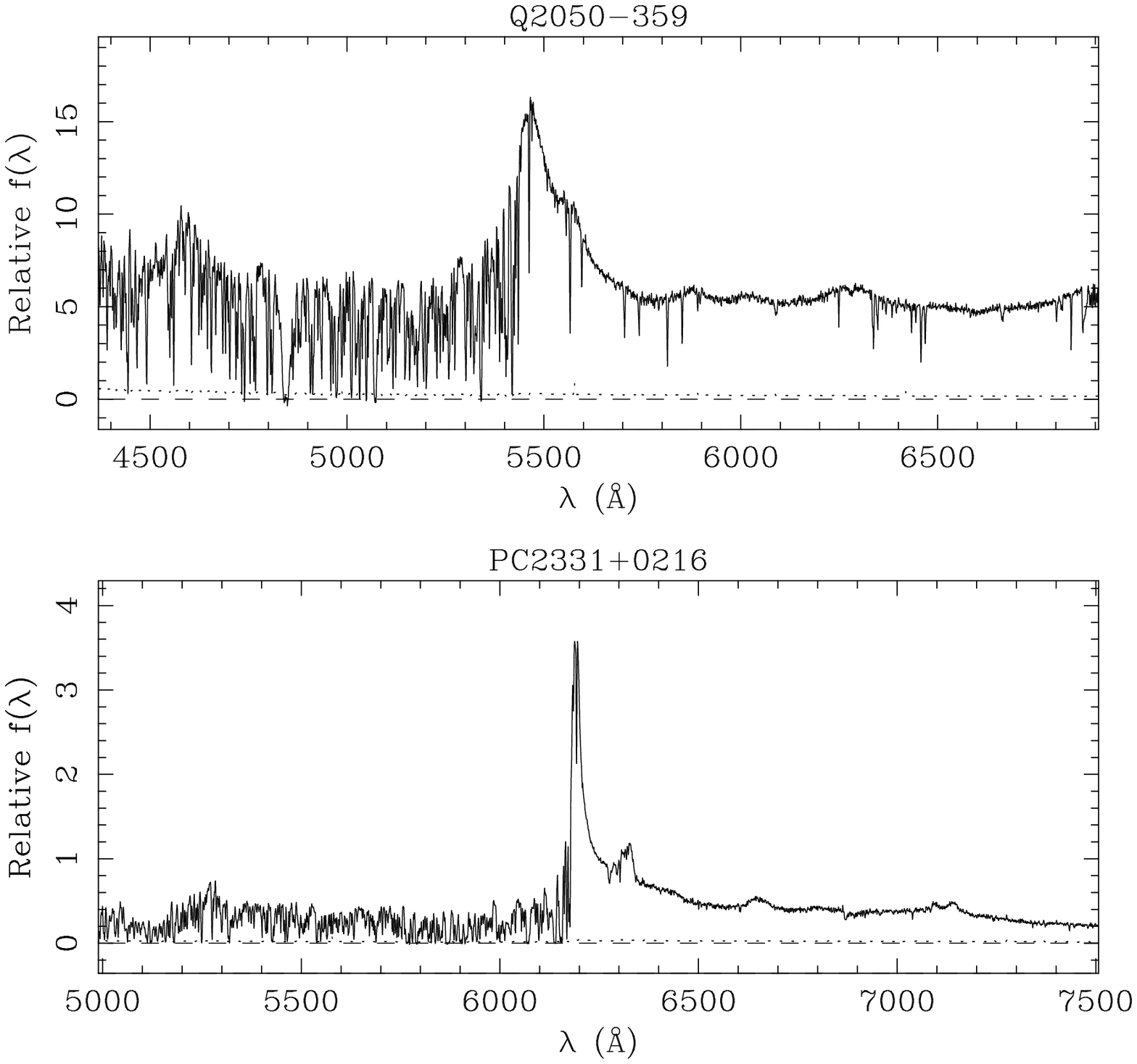}
 
\figcaption{All of the spectra taken in the new LRIS
survey, plus Q0201$+$1120, are shown here. The $1\sigma$ error
arrays are overplotted as dashed lines, though these are difficult
to see in most cases due to the high signal-to-noise ratios
of the spectra. The observations are summarized in table 3.
Expanded versions of the region around the
confirmed damped systems are shown in figure 3.
\label{f_speclris}}
 
\clearpage
%% Figure 3
\plotone{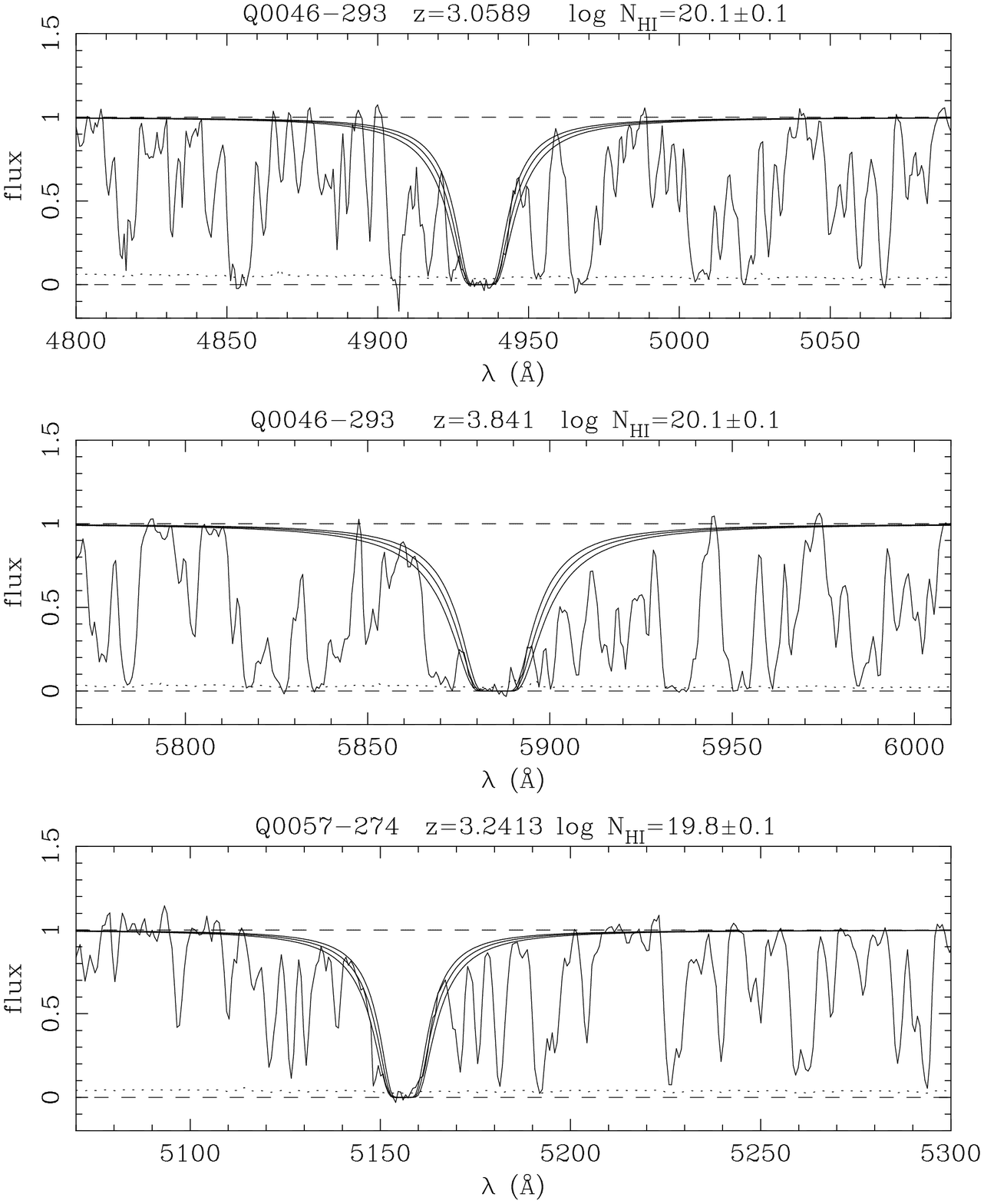}

\plotone{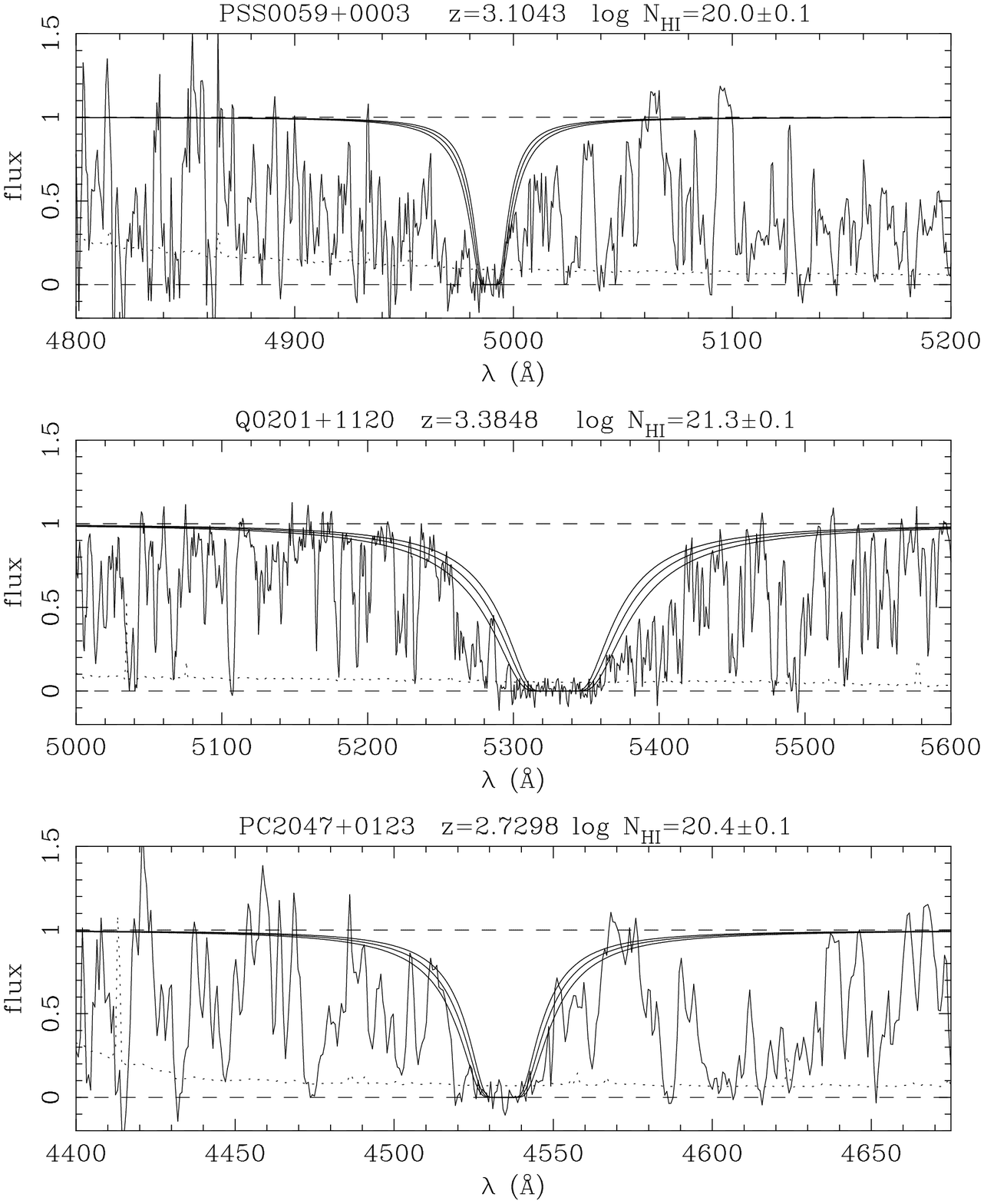}
 
\figcaption{The confirmed
damped \lya systems with log {\nhi} $\ge$ 20.3
cm$^{-2}$ from the new survey are plotted.
Three Voigt profile fits are shown
in order of increasing {\nhi} with the middle
value representing the mean {\nhi} inferred from the optimal fit,
and the other two judged to be $\pm 1 \sigma$ from the mean. The results
are summarized in table 3.
\label{f_specfit}}
 
\clearpage
%% Figure 4
\plotone{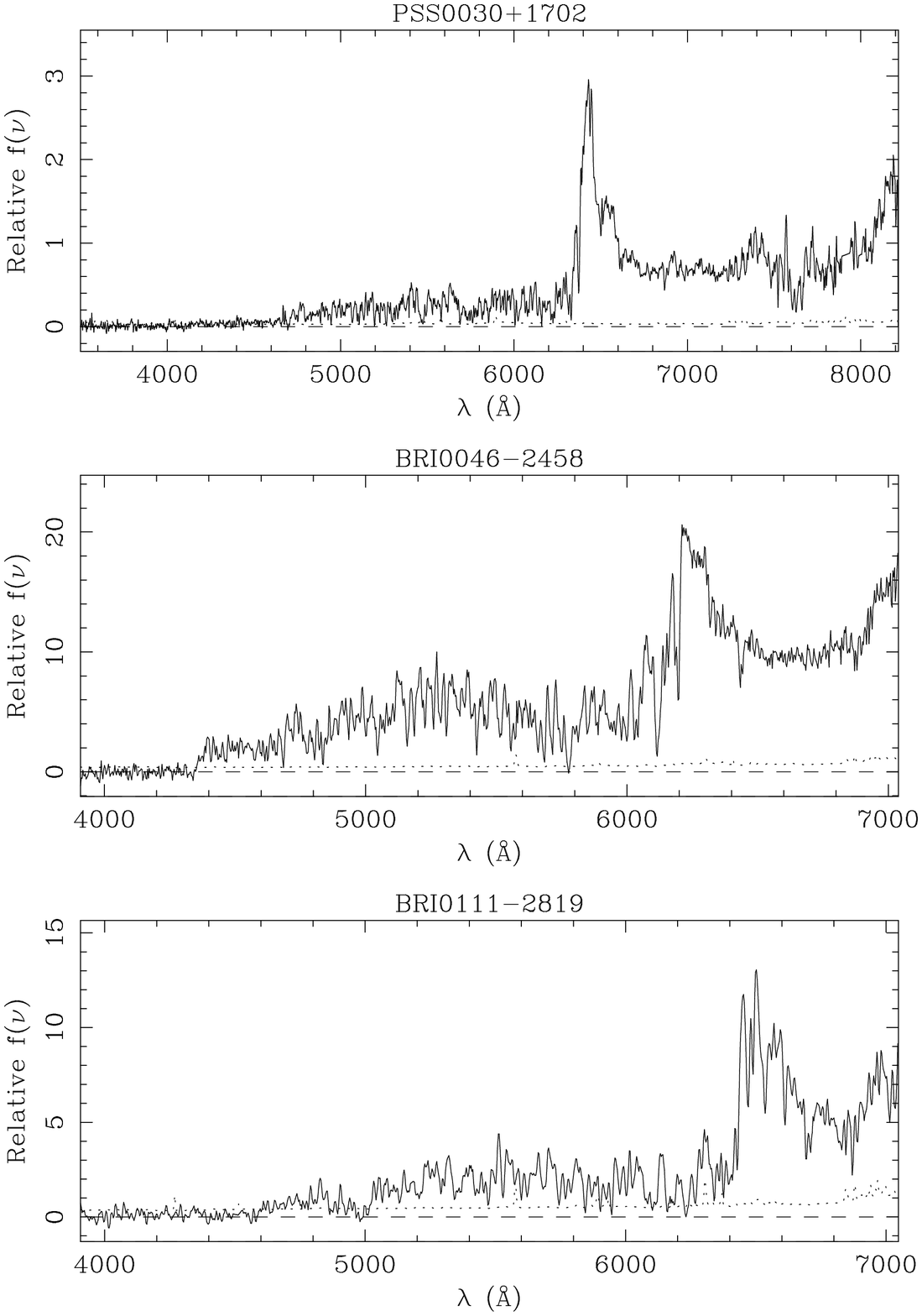}
 
\plotone{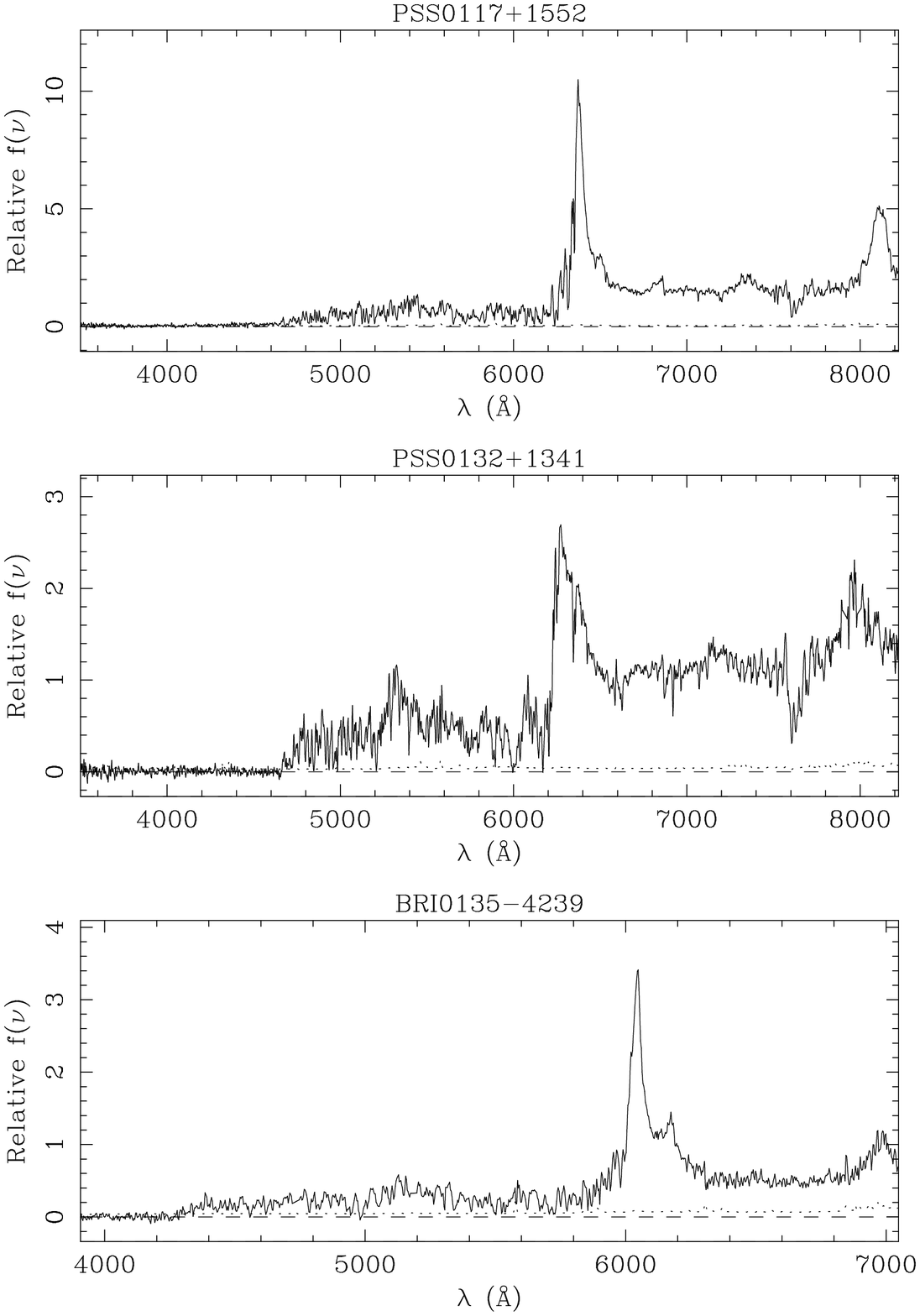}
 
\plotone{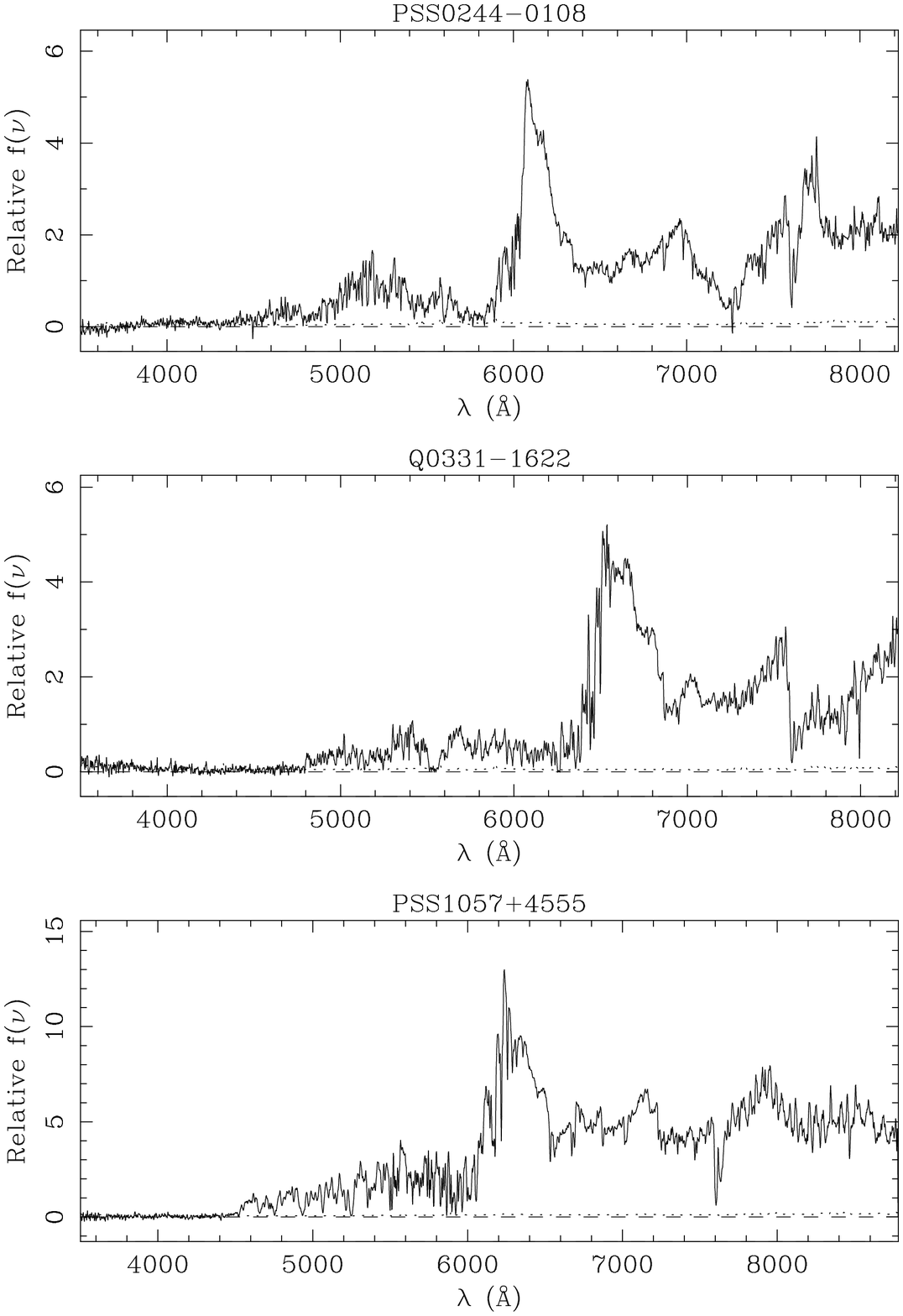}
 
\plotone{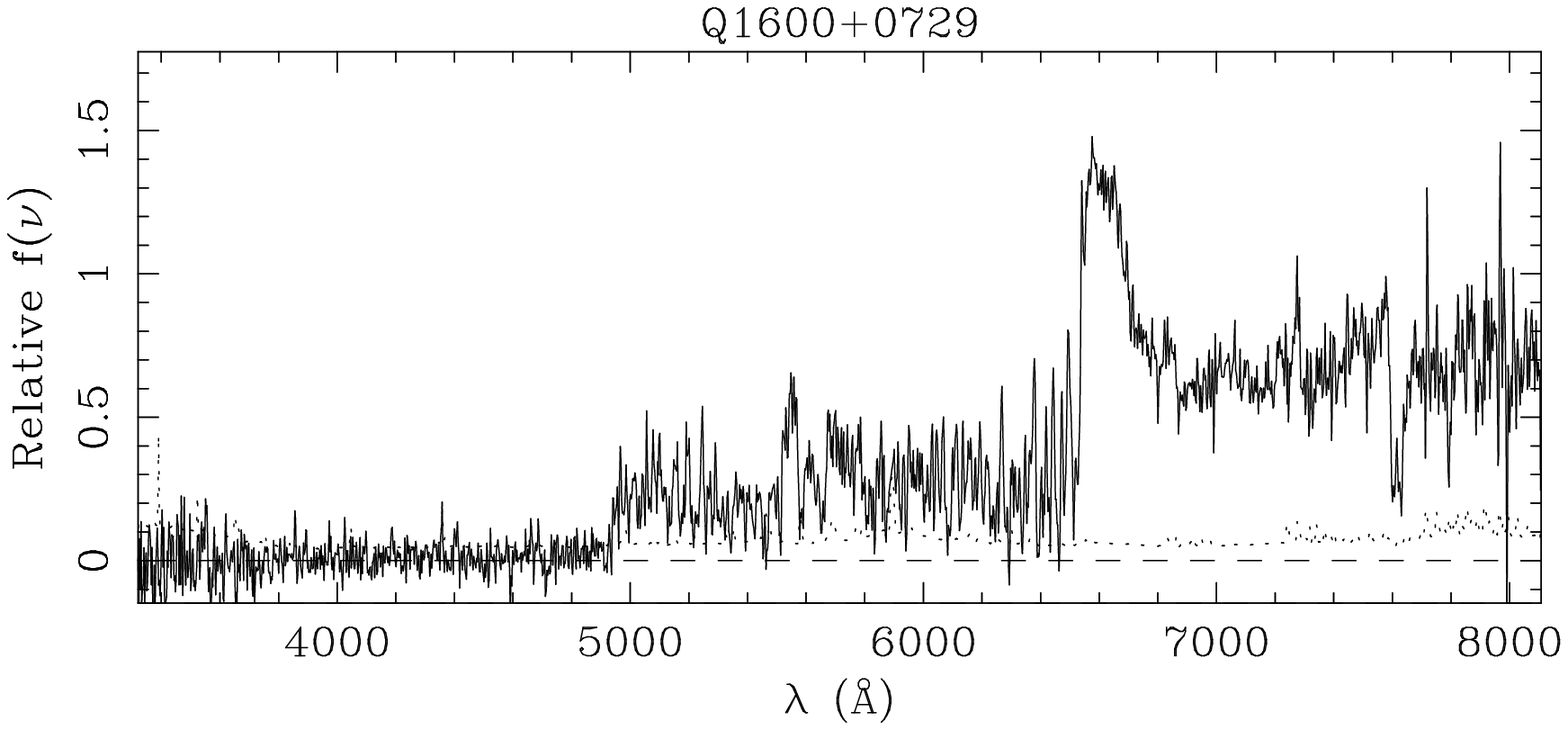}
 
\figcaption{The survey spectra taken at the Lick 3-m and
the Anglo-Australian Telescope are shown. The data are summarized
in table 4.
\label{f_speclick}}
 
%% Figure 5
\clearpage
\plotone{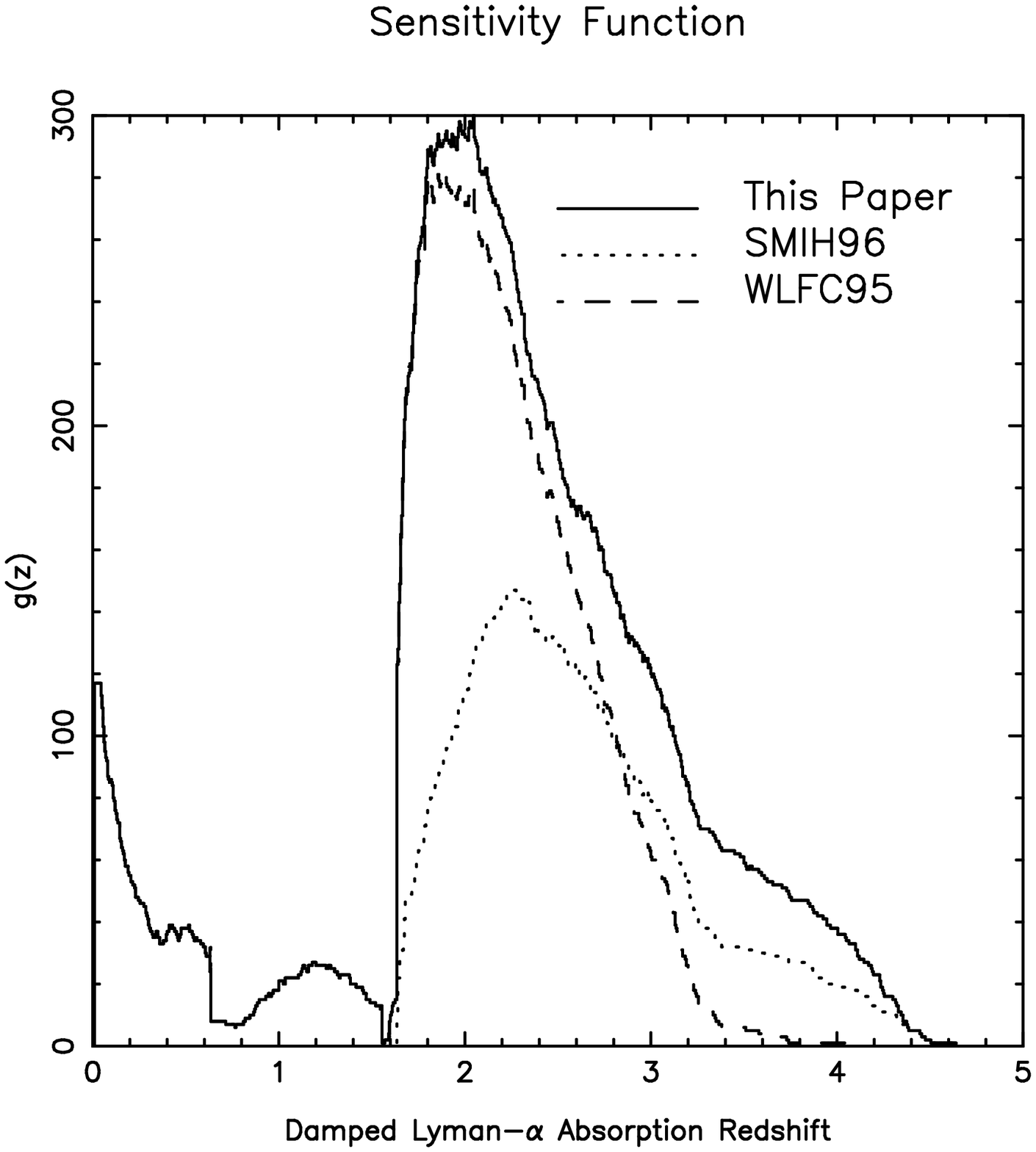}

\plotone{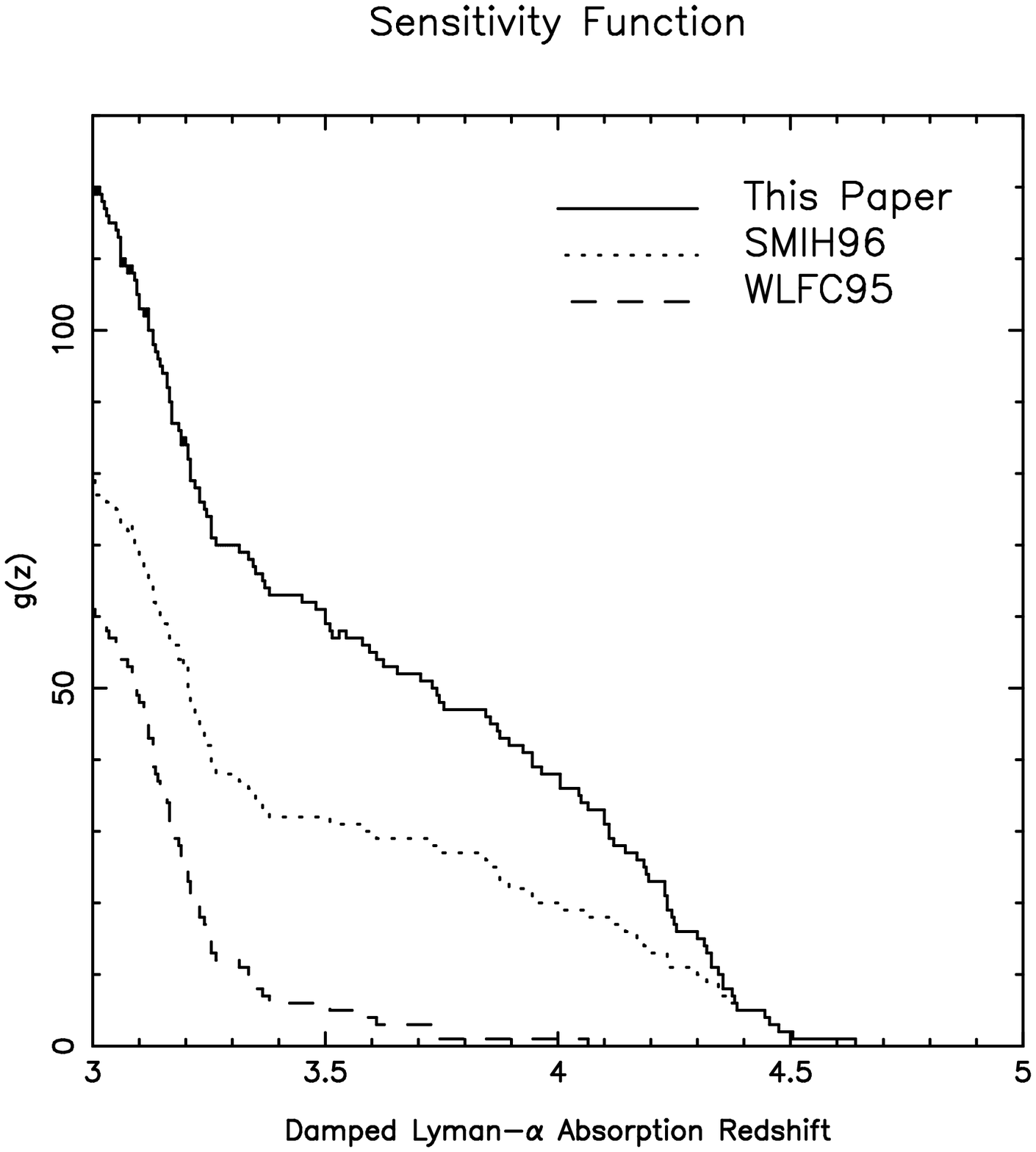}
 
\figcaption{The redshift sensitivity function, $g(z)$, is
plotted for the  combined data sets used in this survey as
a solid line.
It It gives the number of lines of sight at
a given redshift over which damped systems can
be detected at a $> 5\sigma$ level.
The upper panel illustrates how the redshift sensitivity for the entire
statistical sample compares with previous work.
The data included in this paper are shown as a solid line,
the large compilation in the LBQS survey (WLFC95) which
provides the bulk of the data for $z < 3$ is shown as a dashed line, and
the APM survey (SMIH96)
which provided most of the previous data for $z > 3$ is shown as a dotted line.
The lower panel shows the detail for the redshift range $z \ge 3$.
The new data included in this paper nearly double the redshift
path previously surveyed for $z \ge 3$.
\label{f_gz}}
 
\clearpage
%% Figure 6
\plotone{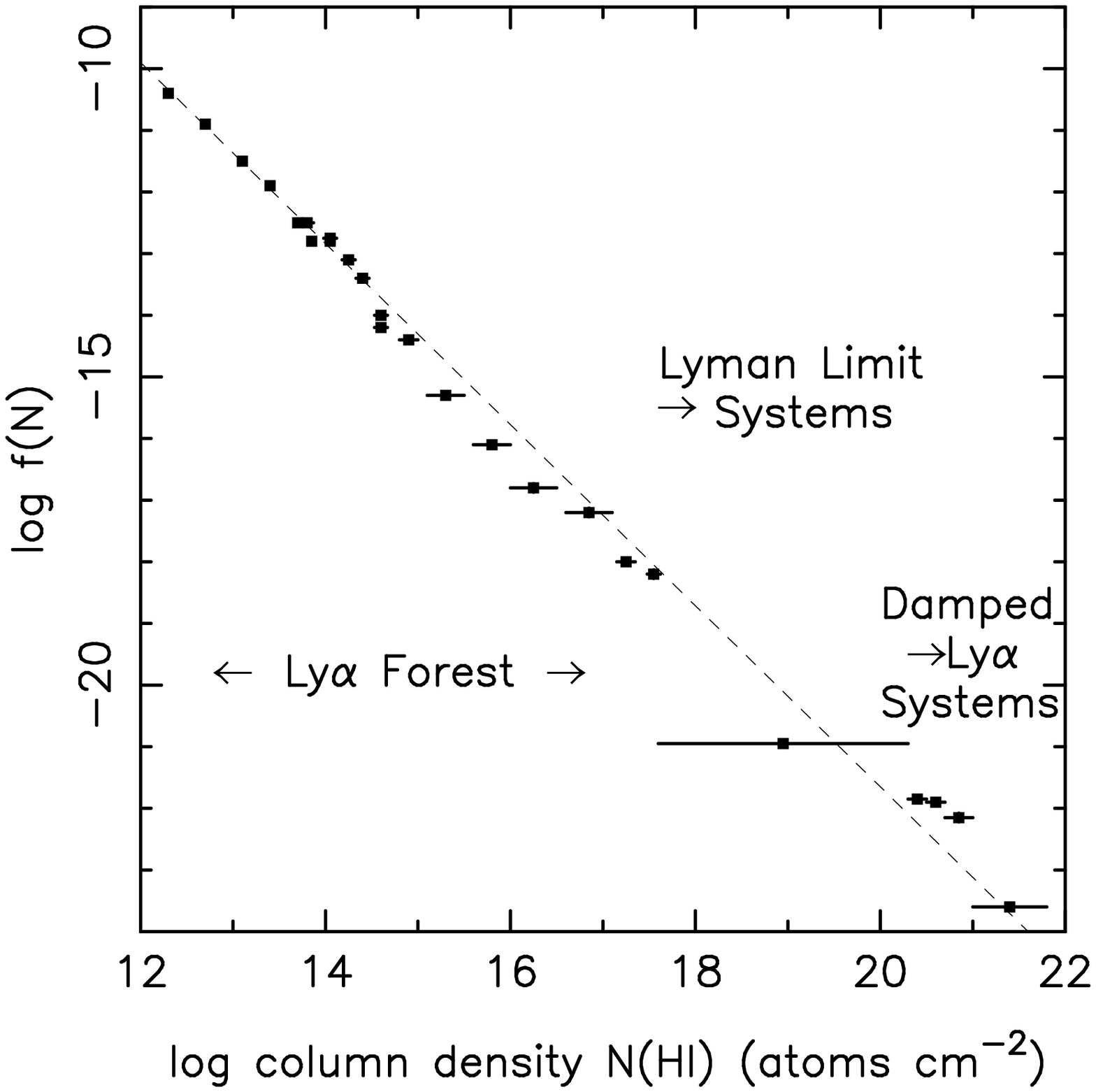}

\figcaption{The column density distribution function of neutral
hydrogen for 12 $\le$ log {\nhi} $\le$ 22 the damped \lya absorbers.
To first order it is fit by a power law, $f(N) \propto N^{-1.46}$,
over 10 orders of magnitude in column density.
\label{f_fnallhi}}
 
%\clearpage
%% Figure 7
\plotone{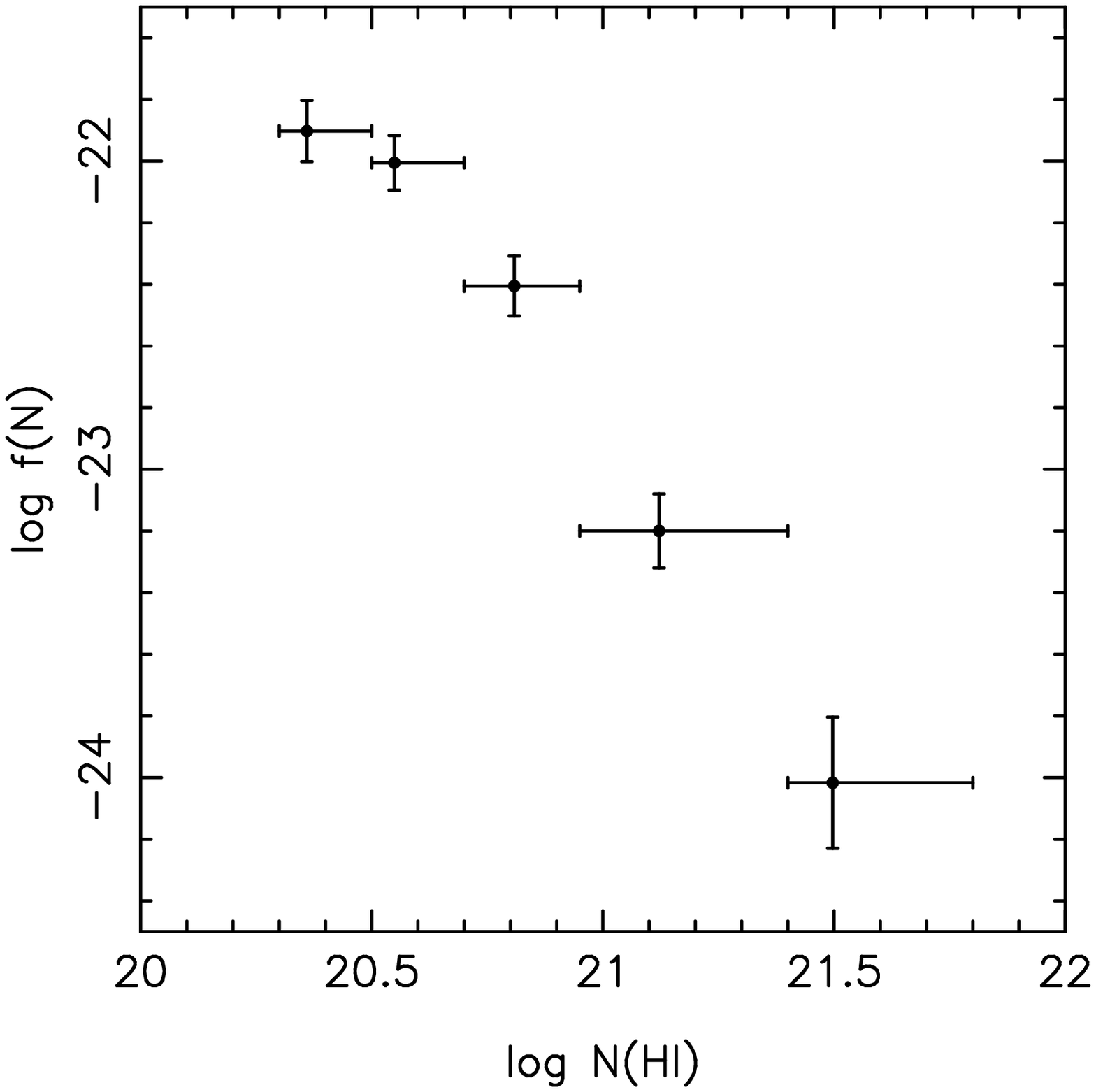}

\figcaption{The differential column density distribution function
of neutral hydrogen determined for the entire
statistical sample of damped \lya absorbers.
\label{f_fnalldiff}}
 
%\clearpage
%% Figure 8

\plottwo{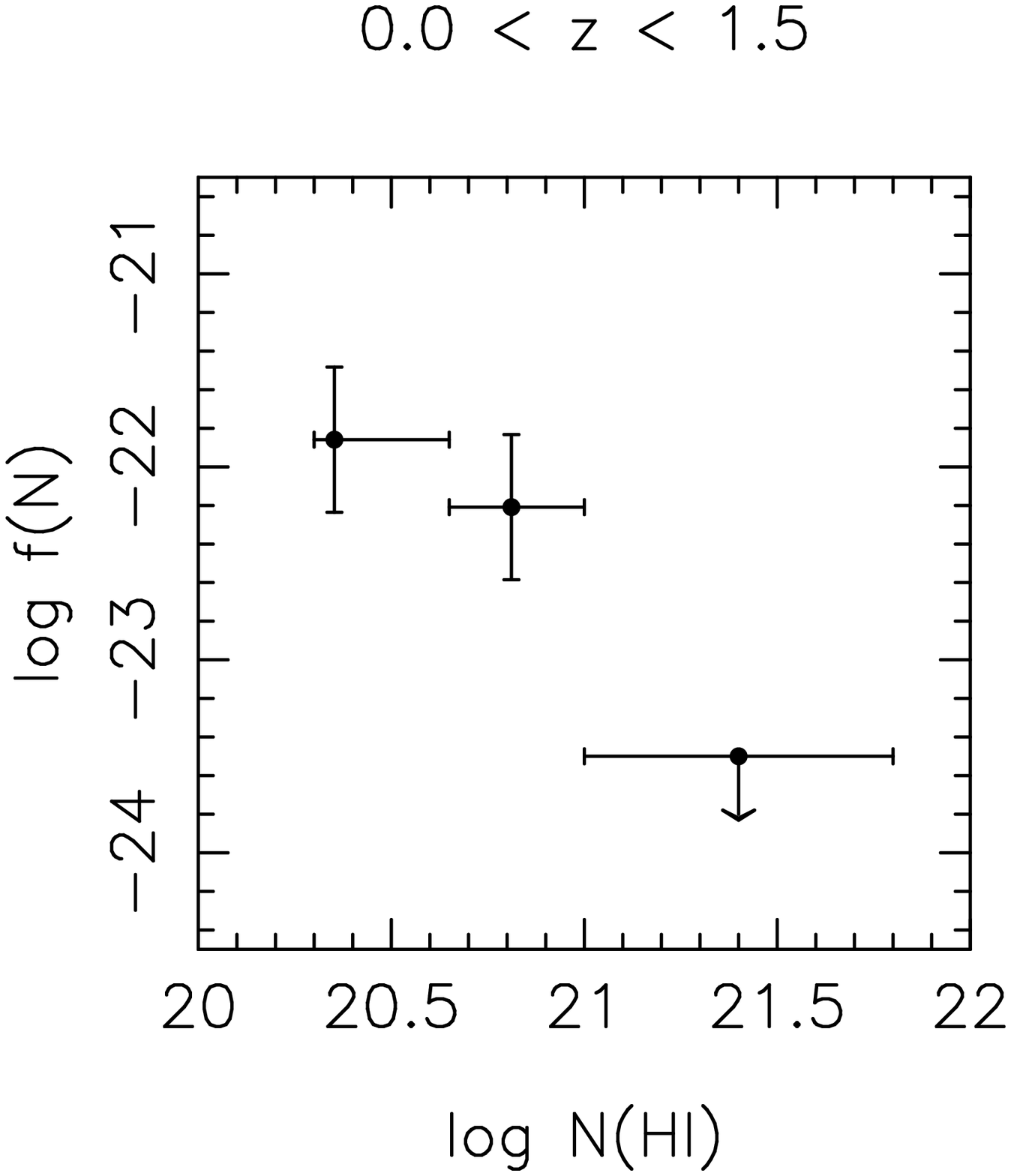}{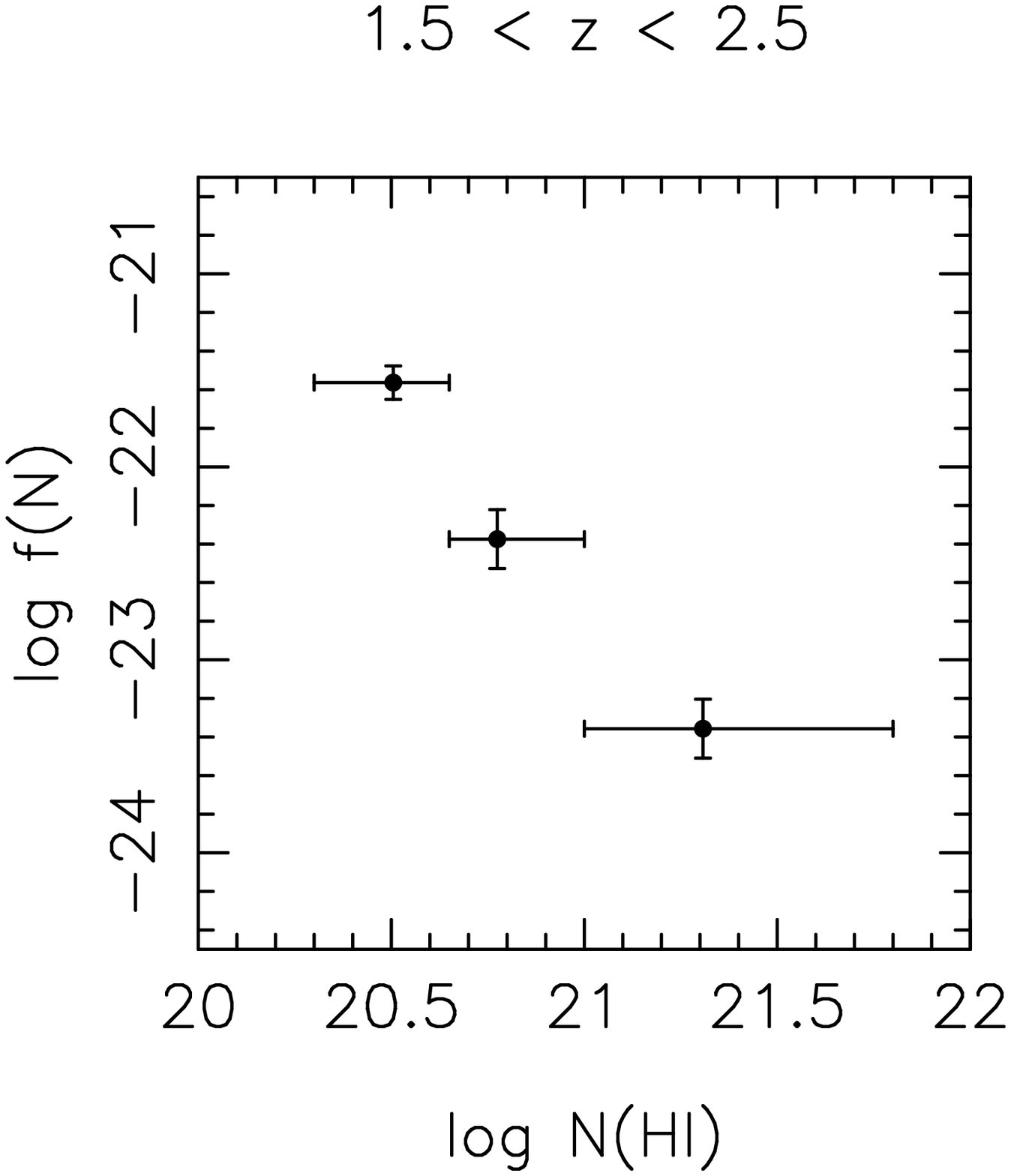}

\plottwo{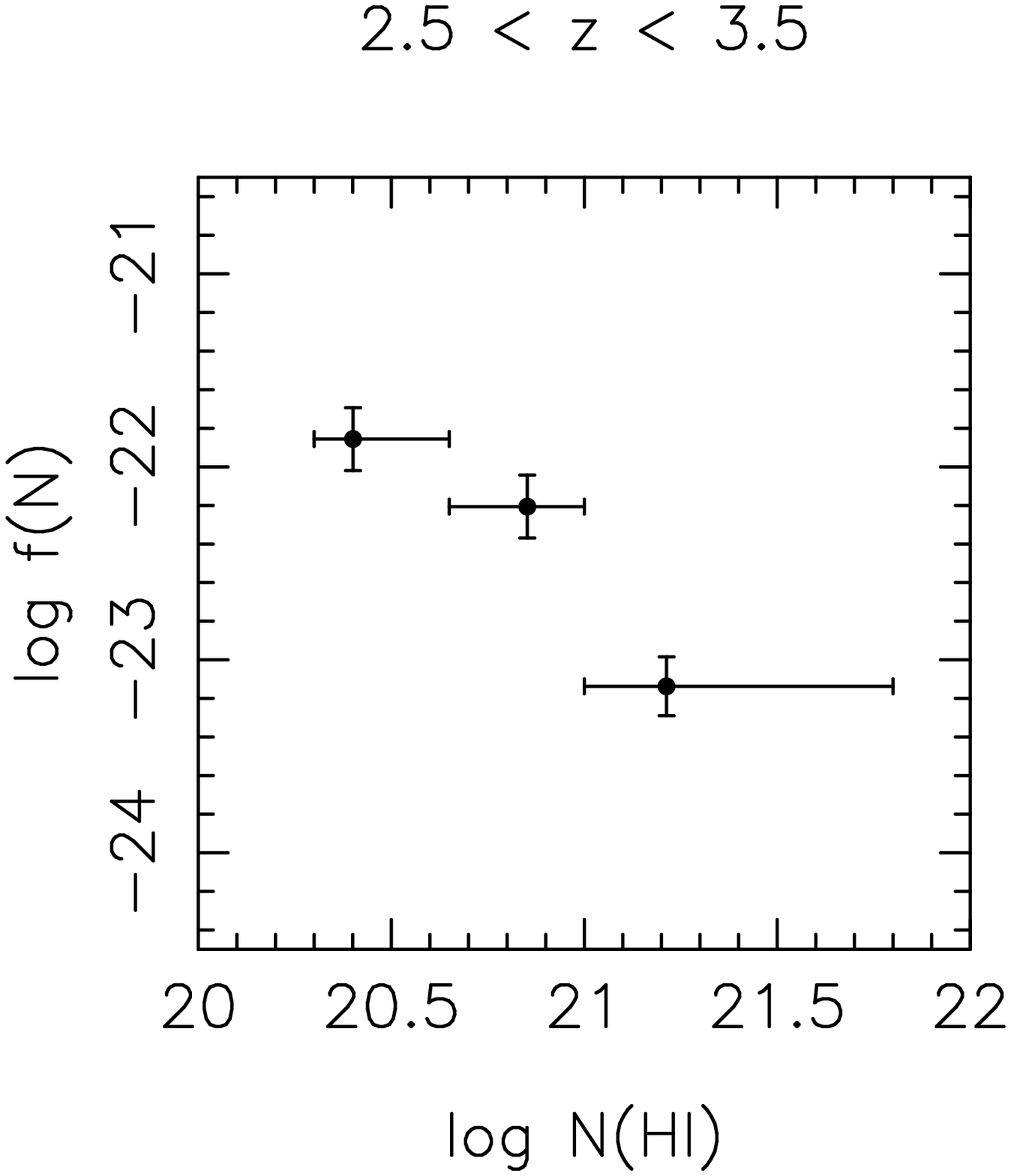}{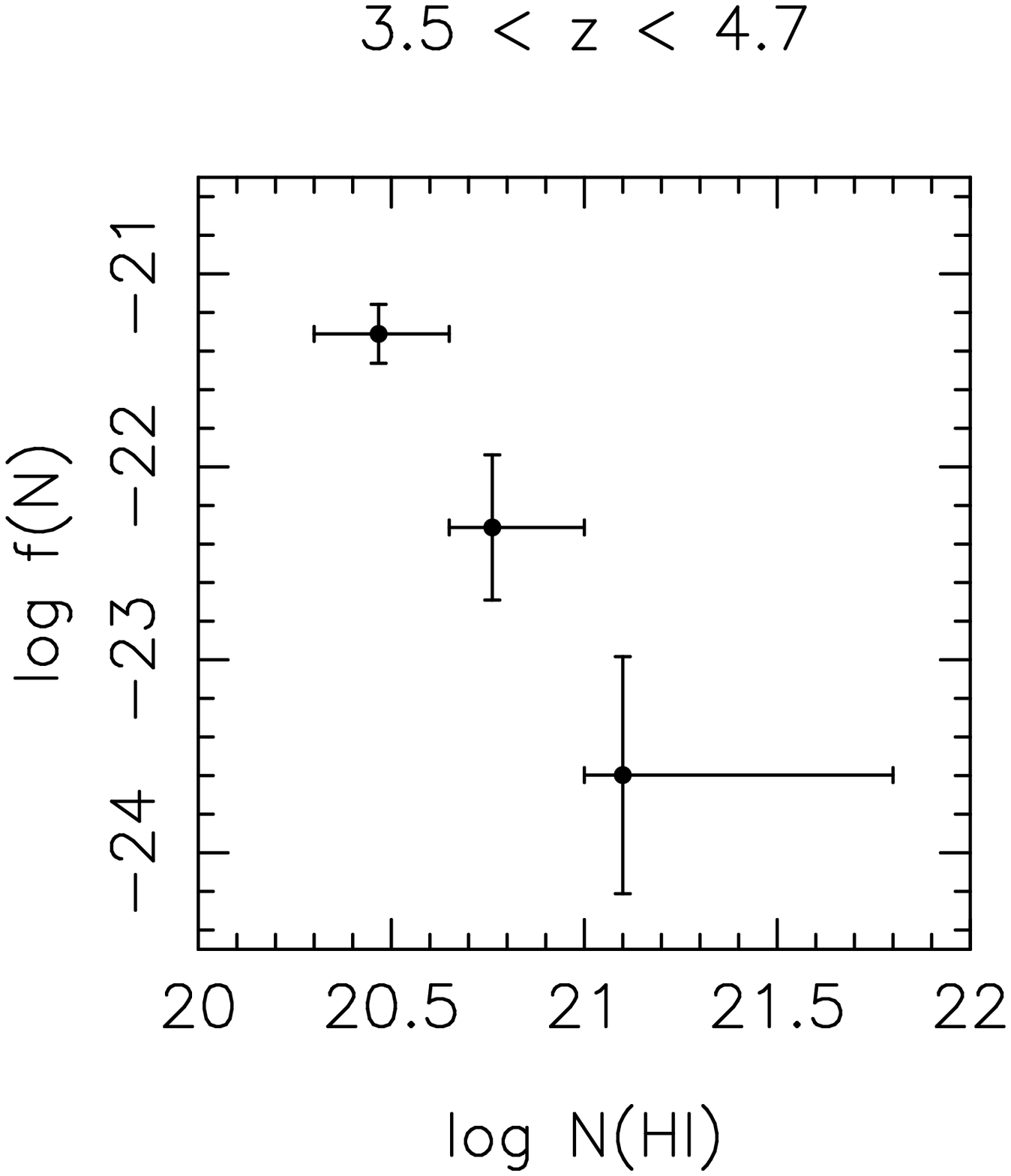}

\figcaption {The differential column density distribution function
of neutral hydrogen
for damped \lya absorbers shown in four redshifts bins:
0.008-1.5, 1.5-2.5, 2.5-3.5, and 3.5-4.7. The distribution is
flattest at z $\approx$ 3 and steepens towards higher and lower
redshifts from that point.  The steepening at high redshift is
due to both a paucity of very high column density systems as well
as an increase in the frequency of lower column density systems.
There is also a steepening of the differential distribution
evident at z $<$ 1.5 but this is not 
statistically significant (see Rao \& Turnshek 2000 
for new results in this redshift range).
\label{f_fnsplitdiff}}
 
\clearpage
%% Figure 9
\plotone{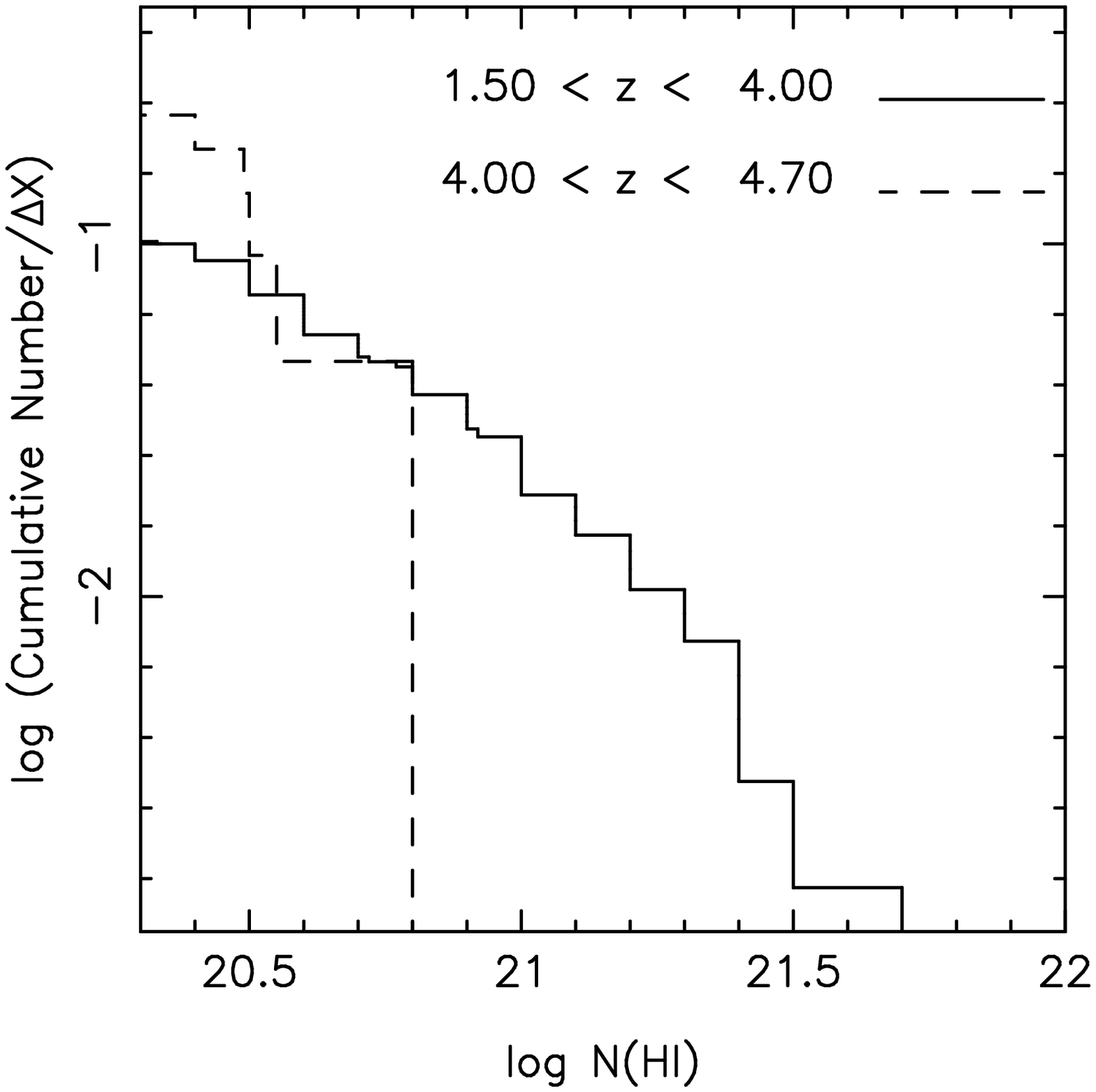}

\figcaption{The log of the cumulative number of
damped \lya absorbers versus log column density
normalized by the redshift path
surveyed, is shown versus log column density
for two redshift ranges.  The solid lines shows the
data for 1.5 $<$ z $<$ 4.0 and the dashed line
shows the data for z $\ge$ 4.0.
A Kolmogorov-Smirnov (K-S) test gives a probability of only 0.006
that the two redshift samples are drawn from the
same distribution.  This is the first time that the
steepening of the column density distribution function
has been shown to be statistically significant.
\label{f_gamcumks}}
 
\vskip 2.0in
%\clearpage
%% Figure 10
\plottwo{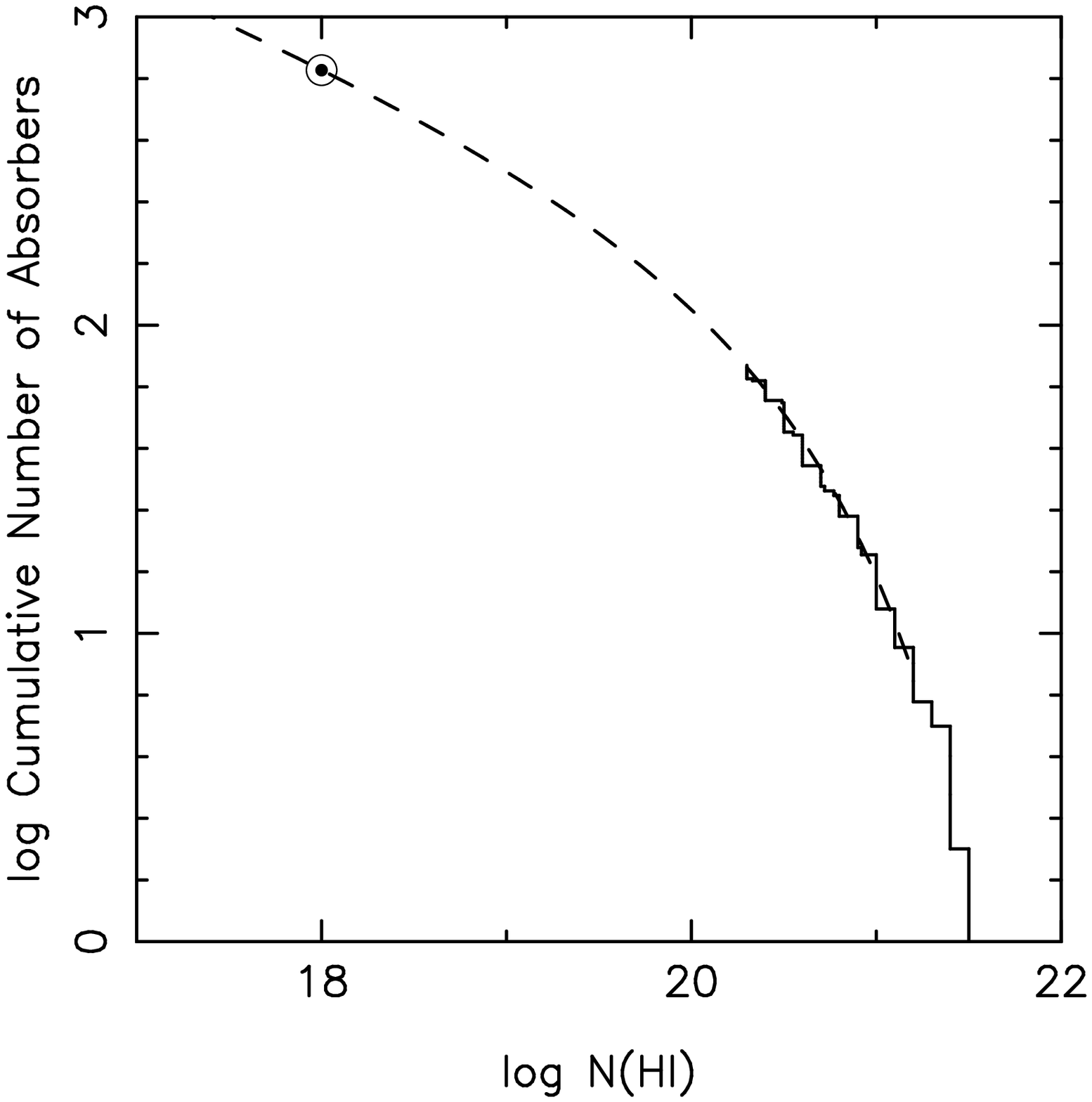}{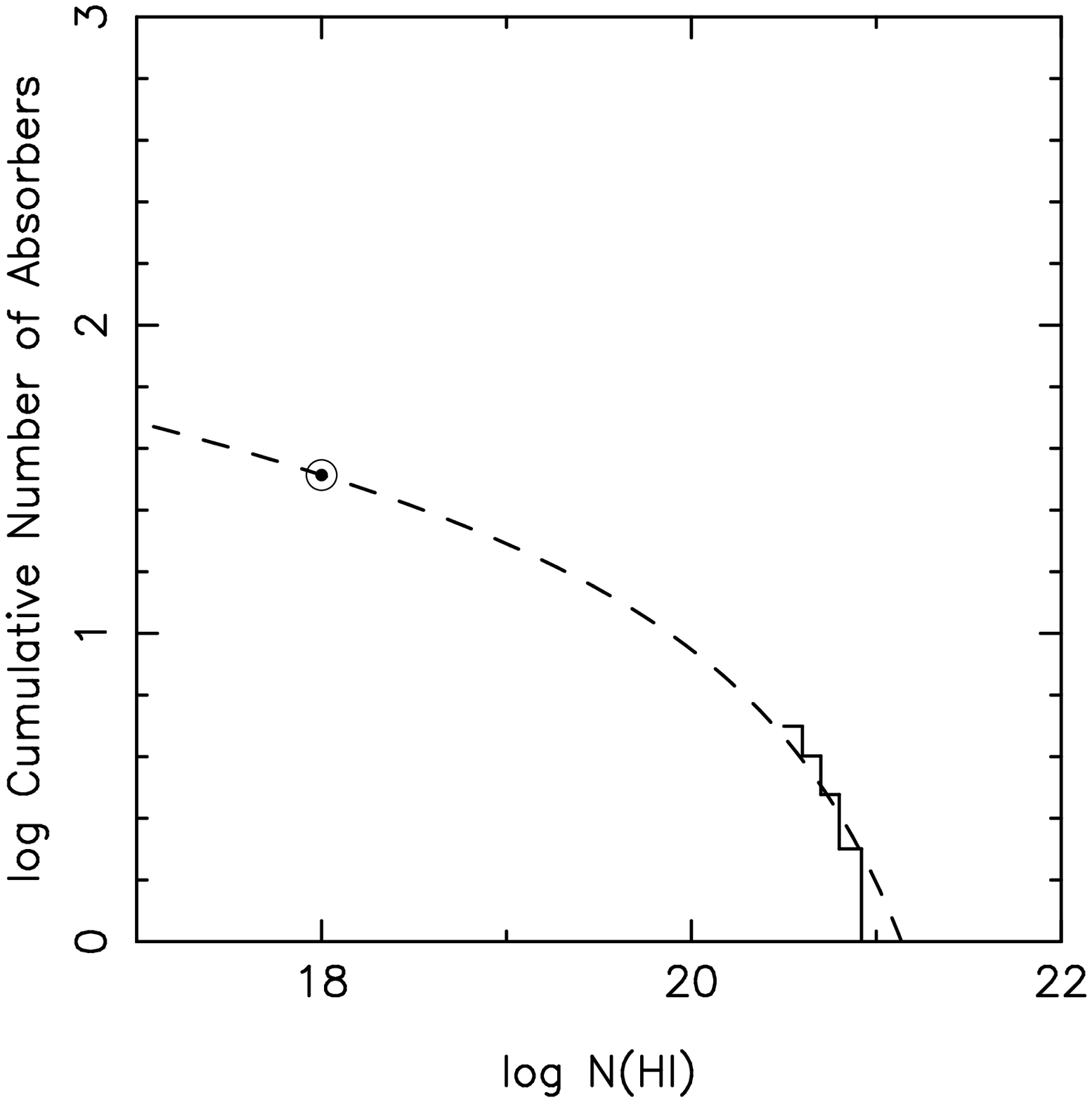}

\plottwo{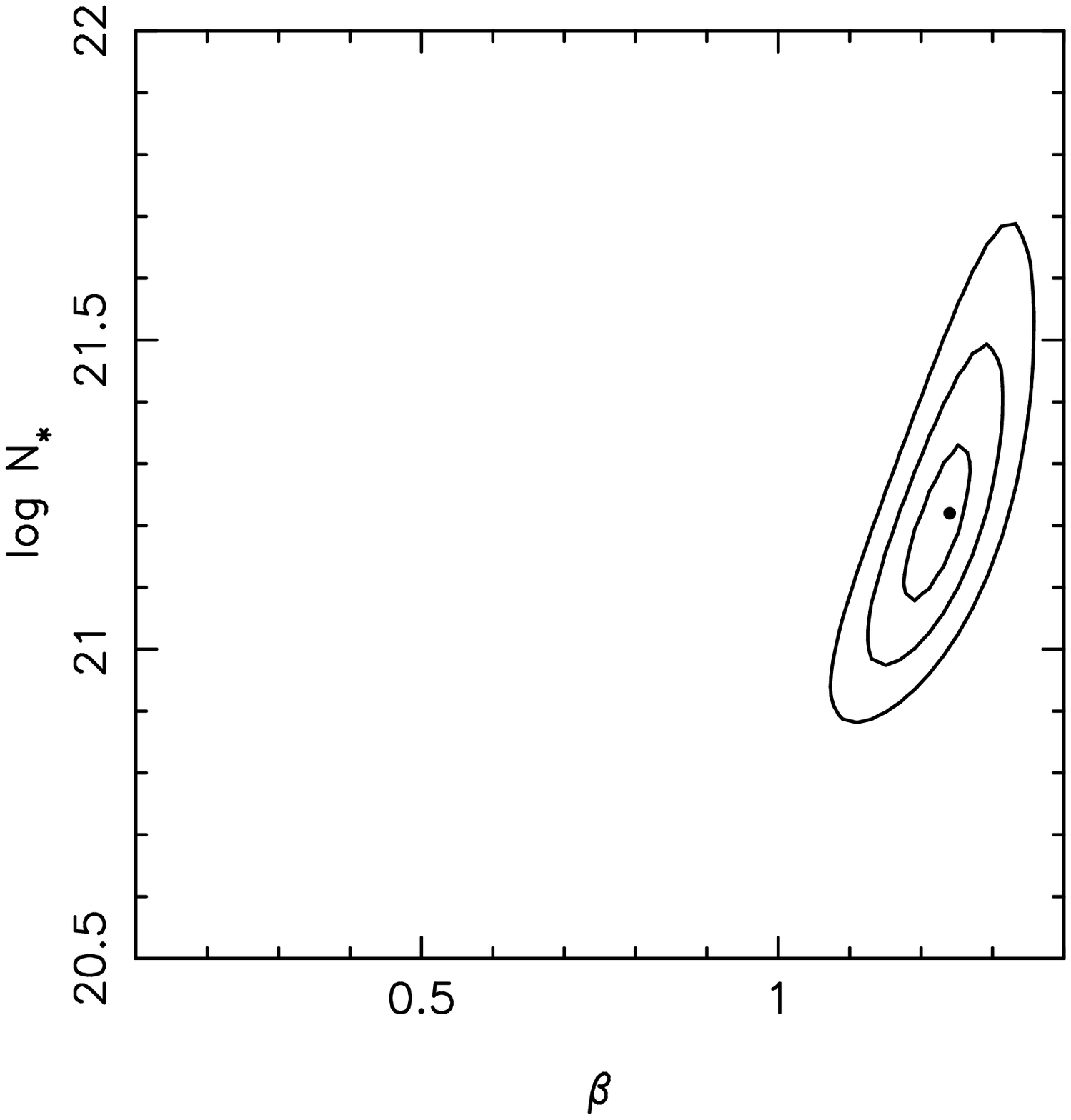}{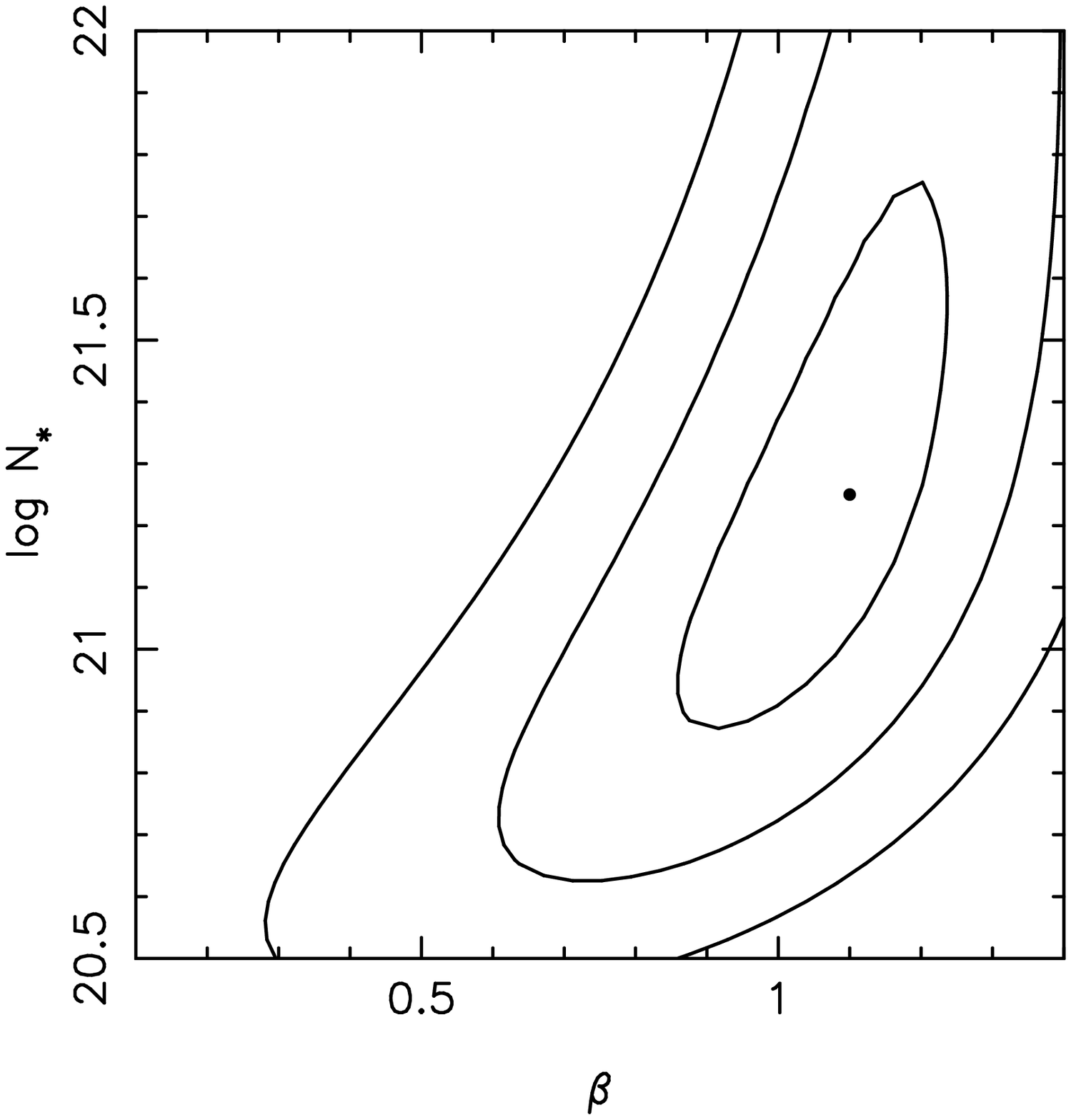}

\figcaption{The left hand plots show 1.5 $<$ z $<$ 4.0 and
the right hand plots show z $\ge$ 4.0. In the top two panels 
we show the log of the cumulative number of
damped \lya absorbers versus log column density, including
a point for the expected number of Lyman limit systems as
a circled star.
The best fits to the column density distribution function
of the form $f(N,z)=( f_* / N_* ) ( N / N_* )^{-\beta} e^{-N/N_*}$
are overplotted using the values in table 8.
In the lower two panels we show the corresponding
$>$68.3\%, $>$95.5\%, and $>$99.7\% confidence contours for
the log-likelihood results when calculating $\beta$, the slope of the
power law portion of the fit, and log N$_{*}$, the knee in 
the distribution.
\label{f_gamcum}}
 
\clearpage
%% Figure 11
\plotone{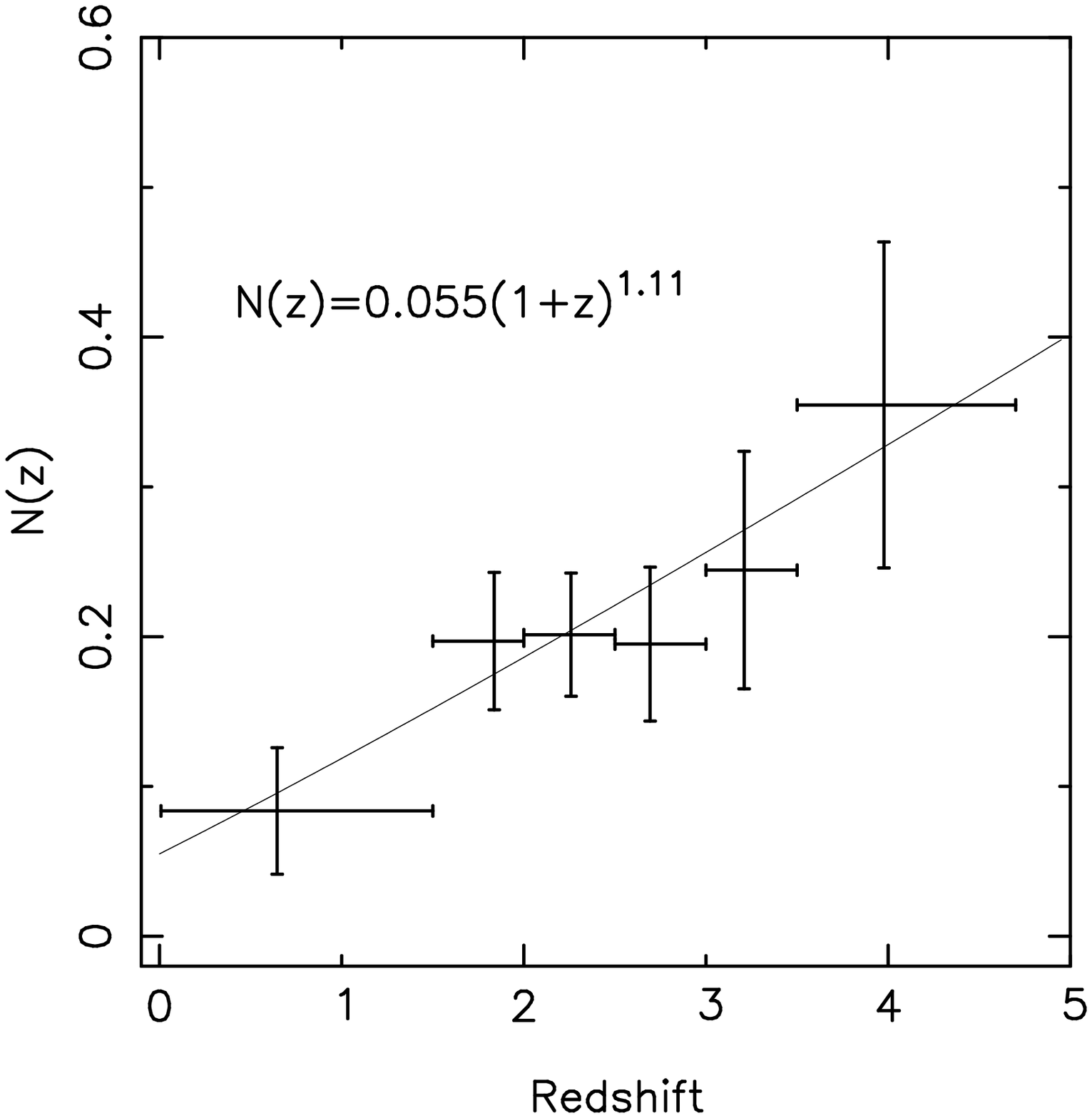}
\figcaption{
The differential distribution in number density of 
absorbers versus redshift for the entire data set is shown. 
It is fit by a
single power law with $N_{0} = 0.055$ and $\gamma = 1.11$, which
would suggest no intrinsic evolution in the product of the
space density and cross-section of the absorbers with redshift,
but the fit is very poorly constrained as shown in figure 12.  
\label{f_dndz_all}}
 
%\clearpage
%% Figure 12
\plotone{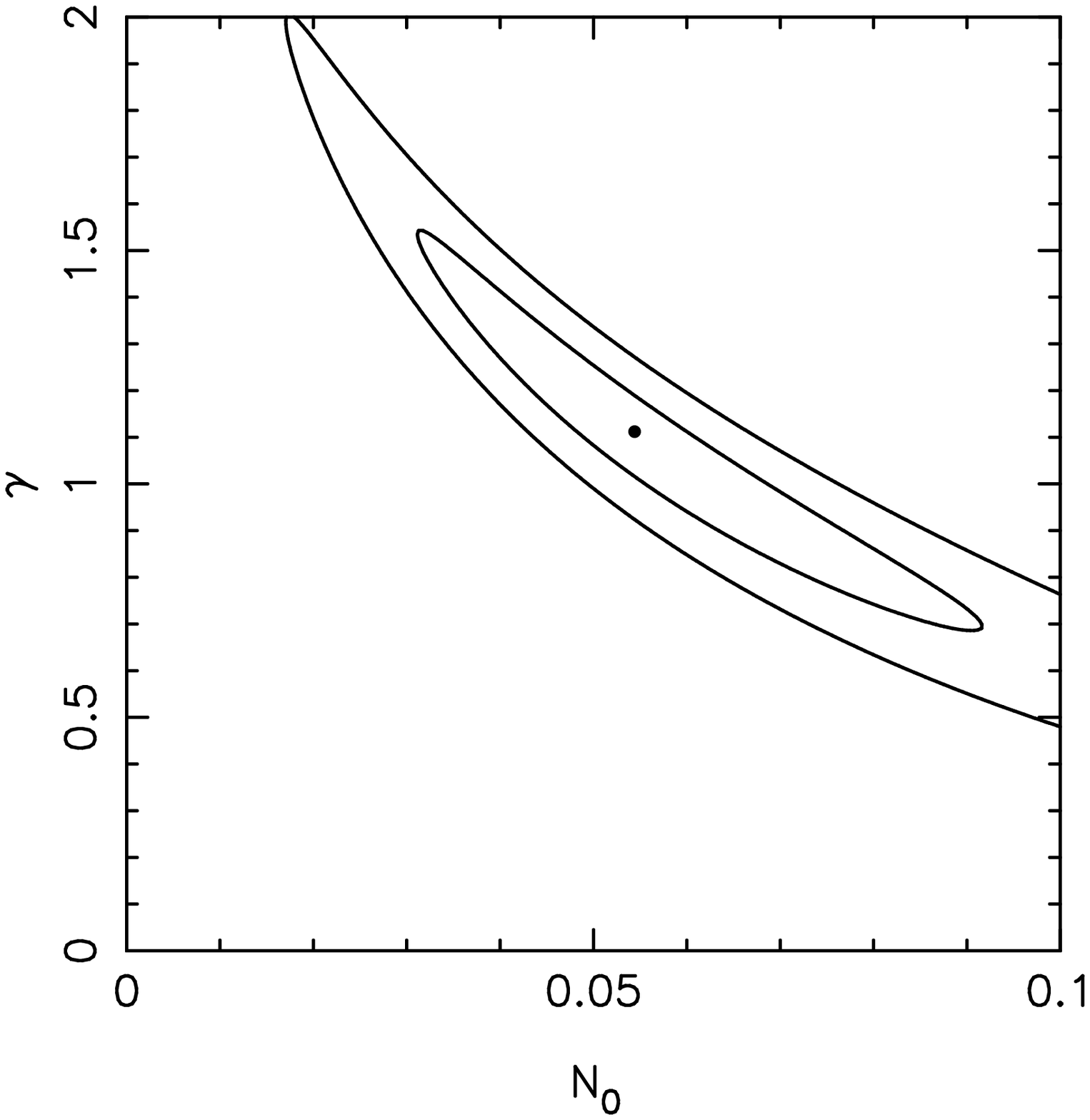}
\figcaption{
The $>$68.3\% and $>$95.5\% confidence contours are 
plotted for the log-likelihood results for $\gamma$ and $N_0$
in fitting the number density per unit redshift with 
a single power law $N(z) = N_{0}(1 + z)^\gamma$.
The fit for a single power law is very poorly constrained
due to differential
evolution in the number density of damped \lya absorbers with different
column densities.  This is illustrated in figure 13. 
\label{f_maxz}}
 
%\clearpage
%% Figure 13
\plotone{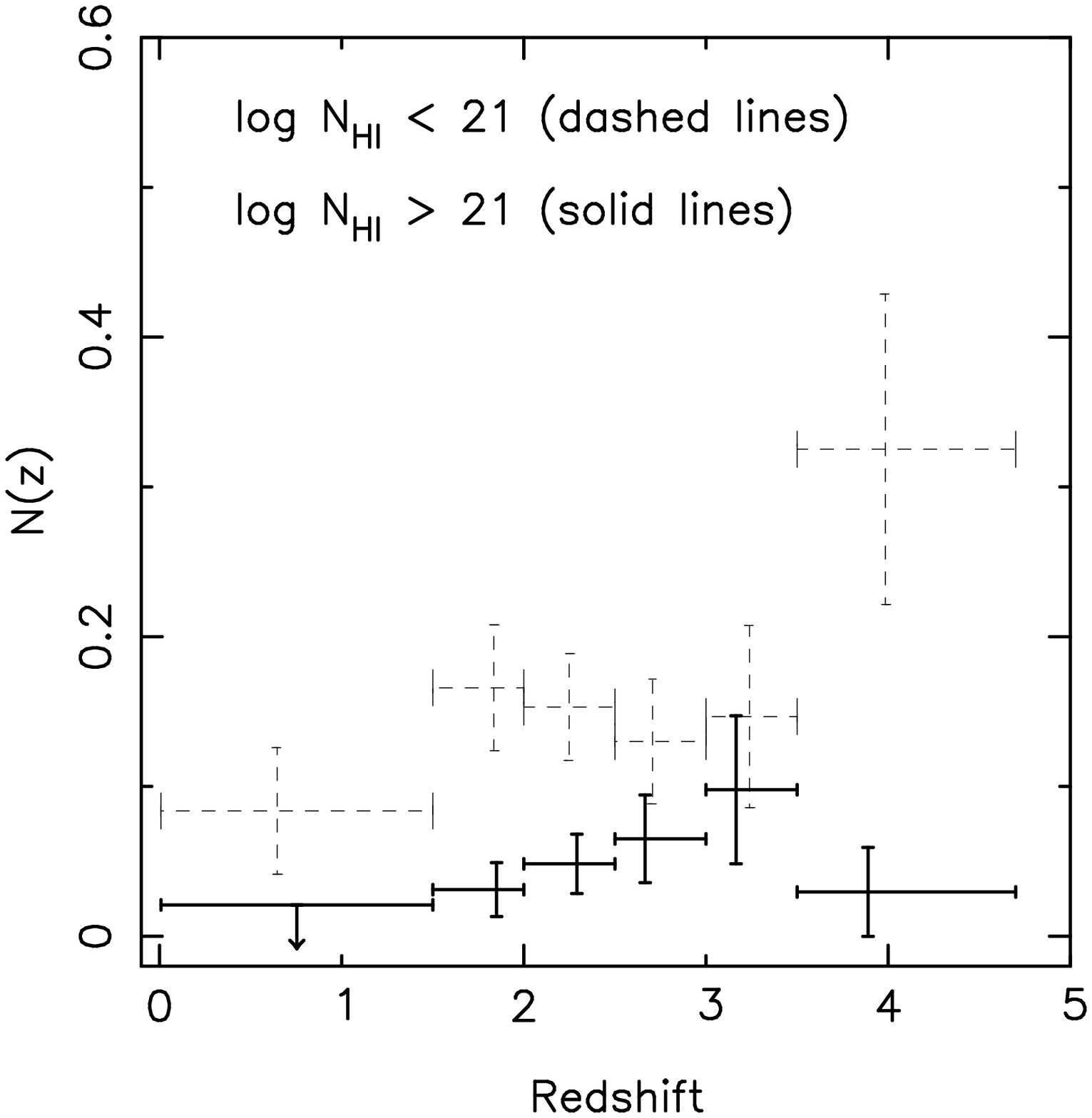}
\figcaption{
The number density per unit redshift is shown for
the same data shown in figure 11, but now split 
into two groups 
at a column density of log {\nhi} = 21.
The number density of systems with log {\nhi} $>$ 21,
shown as solid lines, peaks at z $\approx$ 3.5,
when the Universe is 15-20\% of its present age. These
systems then disappear at a much faster rate from
z$=$3.5 to z$=$0 than does the population of damped absorbers as a whole.
There is a paucity of very high column density systems at
the highest redshifts surveys, 
which was also evidenced in the steepening
of the column density distribution discussed in $\S$~\ref{s_steeper}.
The number density per unit redshift of damped absorbers with 
column densities log {\nhi} $\le$ 21
peaks at z $\approx$ 4, drops at z $\approx$ 3.5 and remains
constant or increases slightly towards z = 1.5.
\label{f_dndz_split}}

\clearpage
%% Figure 14
\plotone{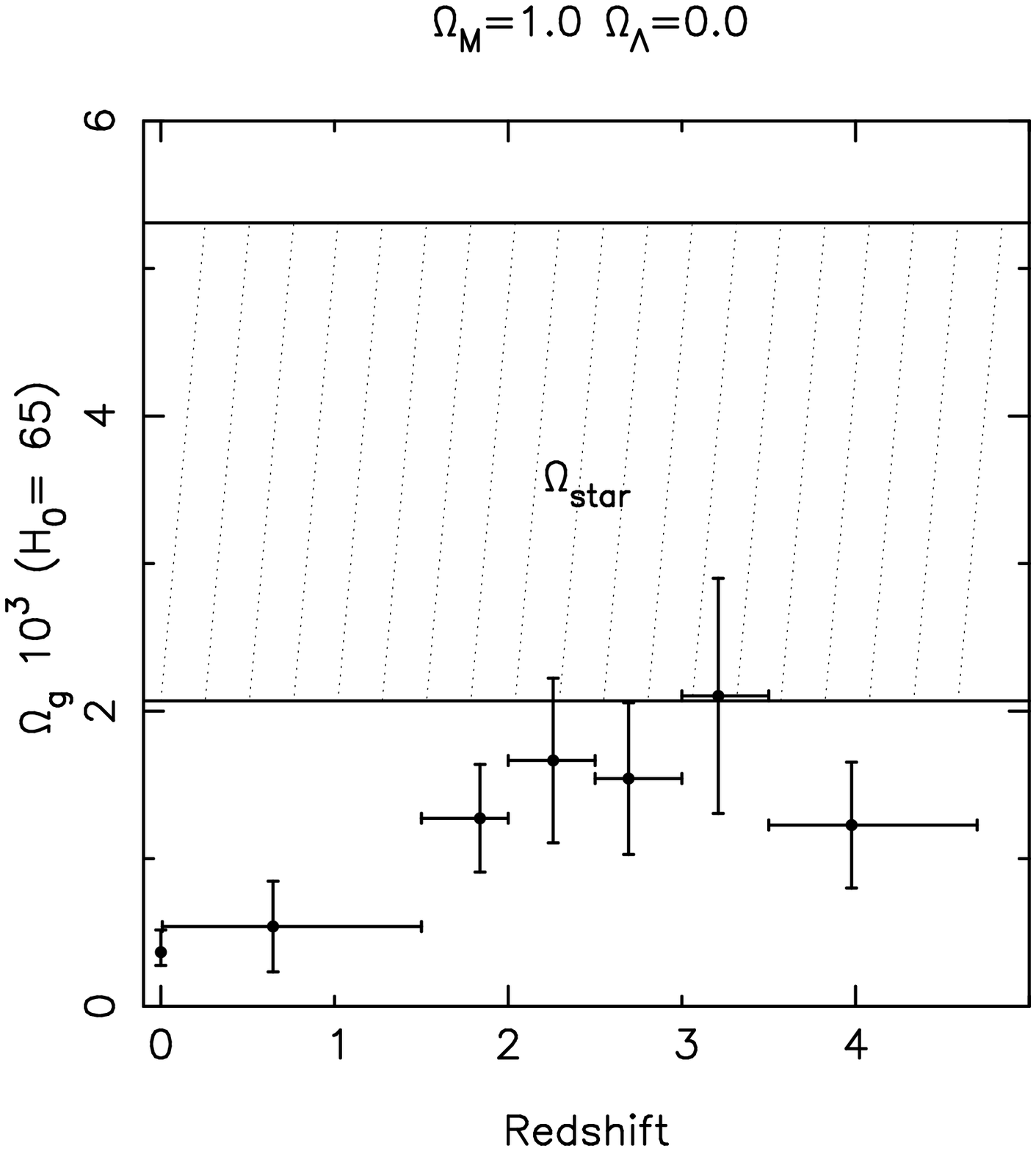}

\plotone{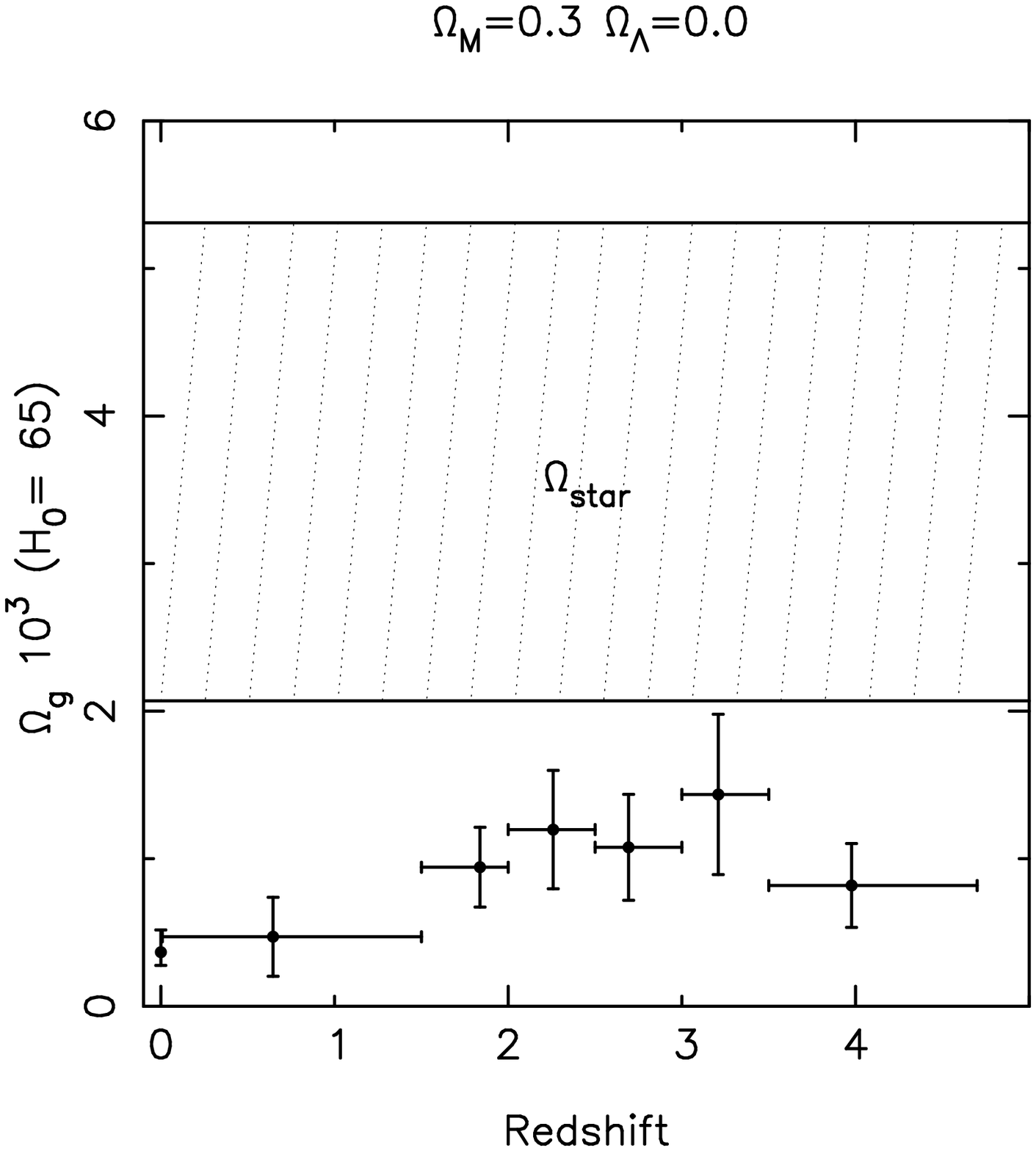}

\plotone{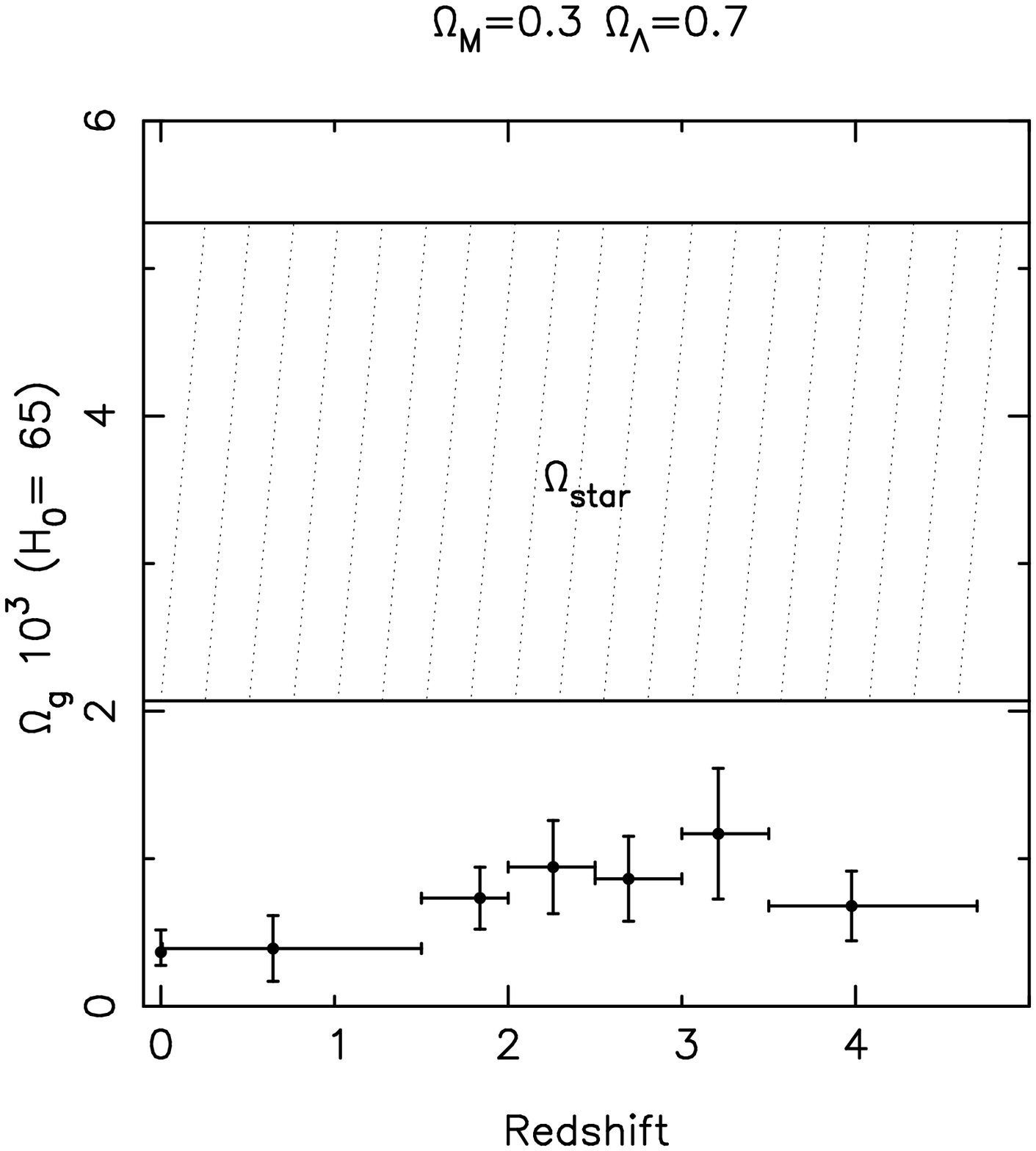}
 
\figcaption{The comoving mass density in neutral gas contributed 
by damped {\lya} absorbers, $\Omega_{g}(z)$, is plotted for three different cosmologies: 
(a) \Wmatter=1, \Wlamb = 0, 
(b) \Wmatter=0.3, \Wlamb = 0, and 
(c) \Wmatter=0.3, \Wlamb = 0.7.
The region {\Wstar} is the $\pm1\sigma$ range for the
mass density in stars in nearby galaxies
(Fukugita, Hogan \& Peebles 1998).  The point at z$=$0 is
the value inferred from 21 cm emission from local galaxies
%(Fall \& Pei 1993; Rao \& Briggs 1993). These results
(Zwaan \etal\ 1997). These results
are tabulated in table 9.
\label{f_omega}}

\clearpage
%% Figure 15
\plotone{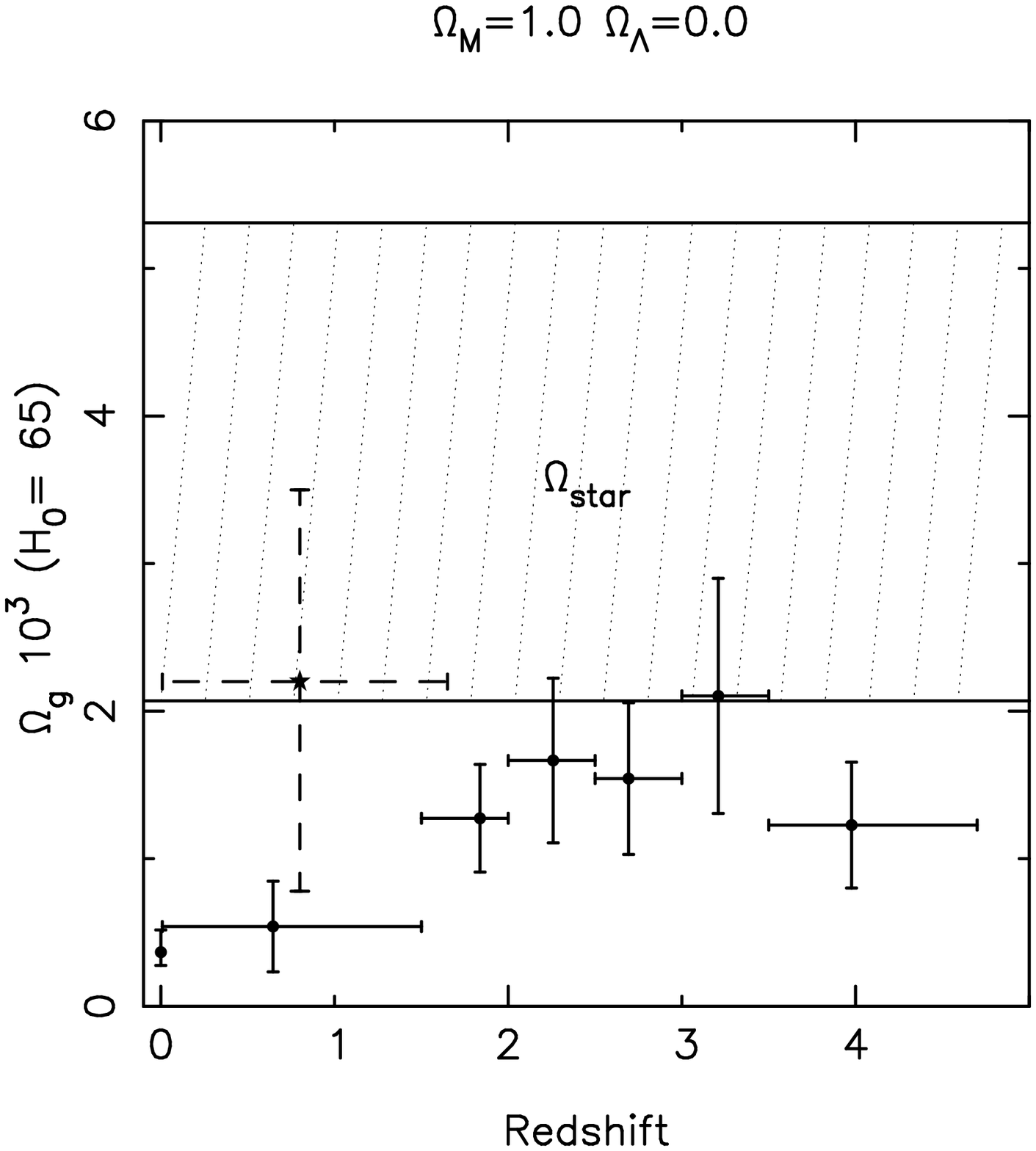}

\figcaption{This is the first panel (\Wmatter=1, \Wlamb = 0) of 
figure 14, showing 
the comoving mass density in neutral gas contributed 
by damped {\lya} absorbers, $\Omega_{g}(z)$,
with the value from Rao \& Turnshek (2000) for $z < 1.65$ plotted 
as dashed lines. The error bars are still very large, but they 
find a substantially higher value of $\Omega_{g}$ at lower redshifts than 
indicated from previous surveys. 
\label{f_omega_RT}}
 
%  TABLE 1 - journal of observations
%  TABLE 2 - followup of known candidates
%  TABLE 3 -  new LRIS survey data
%  TABLE 4 -  new Lick data
%  TABLE 5 -  stat DLA sample
%  TABLE 6 -  stat no-DLA sample
%  TABLE 7 -  redshift path surveyed
%  TABLE 8 -  fit to column density distribution function
%  TABLE 9 -  omega plot values 

\clearpage
%  t1_lris.tex   
%
%%% journal of observations
% 
%\documentstyle[11pt,aaspp4]{article}
%%%%%%%%%%%%%%%%%%
%\documentstyle[apjpt4,myfill]{article}
%\tighten
%\def\etal{et~al.}
%\begin{document}
%%%%%%%%%%%%%%%%%%

\begin{deluxetable}{lllllcrrc} 
\tablenum{1}
\tablewidth{0pc}
%\tabletypesize{\scriptsize}
\scriptsize
\tablecaption{LRIS Journal of Observations\label{t1}}
\tablehead{
\colhead{QSO} & \colhead{z$_{em}$} & \colhead{Date} &
\colhead{V} & \colhead{R} & \colhead{$\lambda_{central}$} &\colhead{Exp} &
\colhead{Grating}& \colhead{Reference} \\
 & &  & &  & \colhead{(\AA)} &\colhead{(secs)} & \colhead{(g/mm)}& 
}
\startdata
BR BB0019$-$1522 & 4.528 & 18 Sep 1995 & &19.0  & 5865 & 3300& 1200 & 1\\
 &  & & & & 7065 & 1200& 1200 \\ 
 &  & & & & 7765 & 1500 & 1200\\ 
Q 0046$-$293&4.014 & 19 Sep 1995 & 19.4&   & 5390 & 2400& 1200 & 2 \\ 
 &  &  & & & 6565 & 1200 & 1200 &  \\
Q 0057$-$274 & 3.52 & 20 Sep 1995 & & 18.73 & 5085 & 1500 & 1200 & 2\\
 &  & & & & 6260 & 600 & 1200 & \\
PSS J0059$+$0003 & 4.16 & 19 Sep 1995 &  & 19.5   & 5190 & 2400& 1200 & 3\\
\enddata
%\tablenotetext{a}{cm$^{-2}$}
\tablerefs{
(1) Storrie-Lombardi {\etal} 1996a
(2) Warren, Hewett \& Osmer 1991 
% Warren, S.J., Hewett, P.C., \& Osmer, P.S. 1991 % \apjs, 76,  23
(3) Kennefick {\etal} 1995
(4) Schneider, Schmidt \& Gunn 1989 % \aj, 98, 1507
(5) McMahon \& Irwin, private communication 
(6) Schneider, Schmidt \& Gunn 1987 % \apjl, 316, L1
(7) White, Kinney \& Becker  1993 % \apj, 407, 456
(8) Sargent, Steidel \& Boksenberg 1989
(9) Schneider, Schmidt \& Gunn 1991 % \aj, 101, 2004
(10) McMahon {\etal} 1994
}
\end{deluxetable}
\vfill

%%%%%%%%%%%%%%%%%%
%\end{document}
%%%%%%%%%%%%%%%%%%

%\clearpage
%  t2_lris.tex
%%% followup of known candidates 
%%%%%%%%%%%%%%%%%%%%%%%%5
%\documentstyle[apjpt4]{article}
%\tighten
%\begin{document}
%\def\lya{Ly$\alpha$}
%\def\etal{et~al.}
%\def\nhi{$N_{HI}$}
%%%%%%%%%%%%%%%%%%%%%%%%5

\begin{deluxetable}{llllcclrlrll}  
\tablewidth{0pc}
%\tabletypesize{\scriptsize}
\scriptsize
\tablenum{2}
\tablecaption{LRIS follow-up of Candidate Damped Lyman-$\alpha$ Absorbers\label{t2}}
\tablehead{ 
& \multicolumn{3}{c}{QSO Properties} & & \multicolumn{4}{c}{Survey Data}
 & & \multicolumn{2}{c}{Follow-up Properties} \\
 \cline{2-4}
 \cline{6-9}
 \cline{11-12}
& \colhead{z$_{em}$} & \colhead{$V$} & \colhead{$R$} &
& \colhead{Ref.} &\colhead{z$_{abs}$}&\colhead{W}& \colhead{log \nhi}&
&\colhead{z$_{abs}$}&\colhead{log \nhi}\\
\colhead{QSO} & & & & & &\colhead{(\AA)} & \colhead{(cm $^{-2}$)} & & &\colhead{(cm $^{-2}$)}
}
%  QSO          z_em    V   R  ref z_ab    W     nhi      z_ab      nhi
\startdata
BR B0019$-$1522 &4.528 & &19.0 & &1& 3.42 & 7.6 & 20.0 & & 3.4370 &20.92$\pm$0.10  \\ 
              &      & &     & &1& 3.98 &12.3 & 20.5 & &\multicolumn{2}{l}{not damped}\\
              &      & &     & &1& 4.28 & 8.0 & 20.1 & &\multicolumn{2}{l}{not damped}\\
BRI B0111$-$2819&4.30  & &18.7 & &2& 3.100&23.4 & 21.0 & & 3.1043 &21.0$\pm$0.1\\
Q 0249$-$2212  &3.202 &18.4 & & &3& 2.831&10.2 & ...  & & 2.8228 &20.2$\pm$0.1  \\
BR B0951$-$0450 &4.369 & &18.9 & &1& 3.84 &24.0 & 21.0 & & 3.8477 &20.6$\pm$0.1  \\
              &      & &     & &1& 4.20 &10.6 & 20.3 & & 4.2028 &20.4$\pm$0.1  \\
BRI B0952$-$0115&4.426 & &18.7 & &1& 4.01 &18.0 & 20.8 &a& 4.0221 &20.2  \\
\enddata
\tablerefs{
(1) Storrie-Lombardi {\etal} 1996a; (2) this paper (3) Sargent, Steidel \& Boksenberg 1989
} 
\end{deluxetable}

%%%%%%%%%%%%%%%%%%%%%%%%5
%\end{document}
%%%%%%%%%%%%%%%%%%%%%%%%5

\clearpage
%  t3_lris.tex
%
%%%  new survey data
%
%\documentstyle[12pt,apjpt4]{article}
%%%%%%%%%%%%%%%%%%%%%
%\documentstyle[apjpt4]{article}
%\tighten
%\def\nhi{\mbox{$N_{\rm HI}$}}
%\begin{document}
%%%%%%%%%%%%%%%%%%%%%
 
\begin{deluxetable}{lllllrr}
\tablewidth{0pc}
%\tabletypesize{\scriptsize}
\scriptsize
\tablenum{3}
\tablecaption{New LRIS Survey Data\label{t3}}
\tablehead{ 
\colhead{QSO}& \colhead{z$_{em}$} 
&\colhead{z$_{min}$}&\colhead{z$_{max}$}&\colhead{z$_{abs}$}&
\colhead{W} &  \colhead{log \nhi} \\ 
&&&&& \colhead{(\AA)}& \colhead{(cm$^{-2}$)} 
}
%  QSO          z_em      z_min z_max z_ab  W  nhi  
\startdata
Q 0046$-$293 &4.014 &  2.882&3.964& 3.0589 &  &  20.1$\pm$0.1 \\
            &      &       &     & 3.8407 &  &(a)19.9$\pm$0.1 \\
            &      &       &     & 3.8422 &  &(b)19.6$\pm$0.1 \\
\multicolumn{4}{r}{components `a' and `b' combined $\rightarrow$}  & 3.841 &&20.1$\pm$0.1  \\
Q 0057$-$274 &3.52  &  2.603&3.475& 3.2413 &  & 19.8$\pm$0.1  \\
PSS J0059$+$0003&4.16  &  2.750&4.108& 3.1043 &  & 20.0$\pm$0.1 \\
PC 0104$+$0215 &4.171 &  2.881&4.119&  &  &      \\
PC 0131$+$0120 &3.792 &  3.116&3.744&  &  &      \\
\enddata
\tablecomments{The confirmed systems have uncertainties quoted
for the column density measurements.  The candidate systems 
do not. Previous experience shows that the system in 
PSS J1443$+$2724 is highly likely to be damped, and the 
candidates in PSS J1435$+$3057 are highly unlikely to be damped.  
All the candidate and confirmed systems with 
log {\nhi}$\ge$20.3 are included in the statistical sample
in the analysis that follows, expect for Q 0201$+$1120.
It was observed because of a known damped absorber, not in 
an unbiased survey.}
\tablenotetext{\dagger}{The emission redshifts for these QSOs have 
been determined from the mean redshifts of the OI 1302, SiIV$+$OIV 1400,
and CIV 1549 emission lines observed in the LRIS spectra.} 
\end{deluxetable}
\vfill

%%%%%%%%%%%%%%%%%%%%%%%
%\end{document}
%%%%%%%%%%%%%%%%%%%%%%%

\clearpage
%  t4_lris.tex
%
%%%  new Lick data
%
%\documentstyle[12pt,apjpt4]{article}
%%%%%%%%%%%%%%%%%%%%%%%%%%%%
%\documentstyle[apjpt4]{article}
%\tighten
%\def\nhi{\mbox{$N_{\rm HI}$}}
%\begin{document}
%%%%%%%%%%%%%%%%%%%%%%%%%%%%
\begin{deluxetable}{lccccclllccc}
\tablewidth{0pc}
%\tabletypesize{\scriptsize}
\scriptsize
\tablenum{4}
\tablecaption{New Lick/AAT Survey Data\label{t4}}
\tablehead{ 
\colhead{QSO}& \colhead{Ref.} & \colhead{R} & Where & \colhead{Date} & \colhead{Exp} 
&\colhead{z$_{em}$} &\colhead{z$_{min}$}&\colhead{z$_{max}$}&\colhead{z$_{abs}$}&
\colhead{W} &  \colhead{log \nhi} \\ 
&&&&&\colhead{(Seconds)}&&&& &\colhead{(\AA)}& \colhead{(cm$^{-2}$)} 
}
%  QSO         ref R Where When  Exp z_em      z_min z_max z_ab  W  nhi  
\startdata
PSS J0030$+$1702& 1 &19.3&Lick & 24 Oct 95 &10800&4.282 & 2.763&4.229&  &  &      \\
BRI B0046$-$2458 &2& 18.9 &AAT &27 Aug 95 & 3600 &4.15 & 2.575 &4.099 &   &   \\ 
BRI B0111$-$2819 &2& 18.7 &AAT &27 Aug 95 & 3295 &4.30 & 2.908 &4.247 &3.100 &23.4&21.0\tablenotemark{b} \\ 
PSS J0117$+$1552& 1 &18.6&Lick & 24 Oct 95 &5400 &4.244 &  2.646&4.192&  &  &      \\
PSS J0132$+$1341& 1 &19.3&Lick & 24 Oct 95 &7200 &4.147 &  2.844&4.096& 3.93 & 10.5 & 20.3 \\
BRI B0135$-$4239 &2& 18.5 &AAT &27 Aug 95 & 4200 &3.97 & 2.575 &3.920 &   &    \\
PSS J0244$-$0108\tablenotemark{a}& 1 &19.0&Lick &24 Oct 95  &3300 &4.00 &       &     & & & \\ 
BR B0331$-$1622  & 2 &17.9&Lick & 23 Oct 95 &9000&4.40 &  2.868&4.326& 3.56 & 14.6  & 20.6  \\
\enddata
\tablecomments{The Lick observations were taken at the 120" with the KAST
spectrograph with simultaneous blue and red arm observations.  The 
Anglo-Australian Telescope observations were taken with the RGO spectrograph.  
All are 
at approximately 6\AA\ resolution (FWHM). For the spectra with enough wavelength
coverage to the red to see CIV, the QSO emission redshifts have been determined
from a mean of the OI 1302, SiIV+OIV 1400, and CIV 1549 emission lines.}
\tablenotetext{a}{This QSO exhibits broad absorption 
lines and is excluded from further consideration.} 
\tablenotetext{b}{This damped candidate was subsequently observed with LRIS 
(see table 2 and figure 2).}
\tablerefs{
(1) Kennefick et al. 1995 and J. Kennefick, private communication
(2) M. Irwin, private communication 
}
\end{deluxetable}
\vfill
%%%%%%%%%%%%%%%%%%%%%%
%\end{document}
%%%%%%%%%%%%%%%%%%%%%%

\clearpage
%  t5_lris.tex
%
%%% statistical sample with DLA
%
%\documentstyle[12pt,apjpt4]{article}
%\tighten
%\def\nhi{\mbox{$N_{\rm HI}$}}
%\def\etal{et~al.}
%\begin{document}
%%%%%%%%%%%%%%%%%%%%%%%%%%
\begin{deluxetable}{llllllc}
\tablewidth{0pc}
%\tabletypesize{\scriptsize}
\scriptsize
\tablecaption{Statistical Sample: QSOs with Damped Lyman-alpha Absorbers\label{t5}}
\tablenum{5}
\tablehead{\colhead{QSO} &  \colhead{z$_{em}$} & \colhead{z$_{min}$} & \colhead{z$_{max}$} & \colhead{z$_{abs}$} & \colhead{log \nhi}  &  Refs.\\ &  &  &  &  &  \colhead{(cm$^{-2}$)}  &  }
%  QSO          z_em      z_min z_max z_ab  nhi  ref
\startdata
Q 0000$-$2619 &  4.11  &  2.389  &  4.060  &  3.3901  &  21.4  &  5,13 \\
Q 0010$-$0012 &  2.15  &  1.634  &  2.119  &  2.0233  &  20.8  &  4,10 \\
Q 0013$-$0029 &  2.08  &  1.634  &  2.049  &  1.9730  &  20.7  &  4,11 \\
BR B0019$-$1522  & 4.528  &  2.97   &  4.473  &  3.4370  &  20.92  &  22,1\\
Q 0027$+$0103   &  2.29  &  1.634  &  2.257 &  1.9375  &  20.6  &  4,10 \\
Q 0042$-$2930   &  2.39  &  1.591  &  2.354 &  1.931   &  20.5  &  4 \\
Q 0049$-$2820   &  2.26  &  1.638  &  2.223 &  2.0713  &  20.5  &  4,11 \\
Q 0056$+$0125   &  3.16  &  2.197  &  3.119  &  2.7750  &  21.0  &  4,10 \\
\enddata
%\vspace{1ex}
\tablerefs{See table 6 for reference list.}
\end{deluxetable}
\vfill
%%%%%%%%%%%%%%%%%%%%%%%%%%
%\end{document}

\clearpage
%  t6_lris.tex
%
%%% statistical sample 
%
%\documentstyle[12pt,apjpt4]{article}
%\tighten
%\def\nhi{\mbox{$N_{\rm HI}$}}
%\def\etal{et~al.}
%\begin{document}
%%%%%%%%%%%%%%%%%%%
\begin{deluxetable}{llllc}
\twocolumn
\tablewidth{20pc}
%\tabletypesize{\scriptsize}
\scriptsize
\tablecaption{Statistical Sample: QSOs without Damped Lyman-alpha Absorbers\label{t6}}
\tablenum{6}
\tablehead{ 
\colhead{QSO}& \colhead{z$_{em}$} &\colhead{z$_{min}$}&\colhead{z$_{max}$} & Survey \\
\multicolumn{5}{r}{Reference}
}
%  QSO          z_em      z_min z_max  ref
\startdata
Q 0001$+$0842  &  3.241  &  2.024 & 3.229  &  21\\
Q 0002$+$151  &  1.8990  &  0.4723  &  0.6034  &  3 \\
--           &  --      &  1.1198  &  1.5500  &  3 \\ 
Q 0003$+$158  &  0.4500  &   0.0080  &  0.4355  &  3 \\
MG 0004+1359   &  3.25  &  1.899  &  3.207   &  2 \\ 
Q 0004$+$1711  &  2.898  &  2.002 & 2.851  & 21 \\ 
Q 0006$+$0230   &  2.09  &  1.787  &  2.059  &   4 \\  
Q 0006$+$0200   &  2.35  &  1.634  &  2.317  &   4 \\
\enddata
\tablerefs{
(1) this paper; 
(2) Storrie-Lombardi \& Hook 2000; 
(3) Lanzetta, Wolfe, \& Turnshek 1995; 
(4) WLFC95;
(5) Sargent, Steidel, \& Boksenberg 1989; 
(6) Turnshek {\etal} 1989; 
(7) Wolfe {\etal} 1993; 
(8) Lu {\etal} 1993; 
(9) Lu \& Wolfe; 
(10) Virgilio {\etal} 1995; 
(11) Pettini {\etal} 1994; 
(12) Francis \& Hewett 1993; 
(13) Savaglio {\etal} 1994; 
(14) Sargent, Boksenberg, \& Steidel 1988; 
(15) Black, Chaffee, \& Foltz 1987; 
(16) Wolfe \etal\ 1986;
(17) Wolfe {\etal} 1994; 
(18) Rauch {\etal} 1990; 
(19) Williger {\etal} 1989;
(20) Meyer, Lanzetta \& Wolfe 1995;
(21) Lanzetta {\etal} 1991; 
(22) SMIH96; 
(23) SIM96
}
\end{deluxetable}
\vfill
%%%%%%%%%%%%%%%%%%%%%%
%\end{document}

\clearpage
%  t7_lris.tex
%
%%%  redshift path summary 
%
%%%%%%%%%%%%%%%%%%%%%
%\documentstyle[apjpt4]{article}
%\tighten
%\begin{document}
%%%%%%%%%%%%%%%%%%%%%
\begin{deluxetable}{crrrrr}
\tablewidth{0pc}
%\tabletypesize{\scriptsize}
%\scriptsize
\tablecaption{Redshift Path Surveyed\label{t7}}
\tablenum{7}
\tablehead{
\colhead{Redshift} & \colhead{Survey}& \colhead{$\Delta$z} &
\multicolumn{2}{c}{$\Delta X$} & \colhead{\# of} \\
\colhead{Range} & &&  \multicolumn{2}{c}{q$_0$=0 \ \ \ \ q$_0$=0.5}&\colhead{QSOs} }
\startdata
3.00 - 5.00 &  This Paper &   73.8 & 332.5   & 156.5   &     128 \\
  &    APM &  43.5 & 195.8   &  92.2   &      82 \\
  & LBQS &     13.4 &  56.3   &  27.5   &      61 \\
\multicolumn{6}{l}{ \ } \\
0.01 - 5.00 & This Paper & 418.4 & 1364.2 &747.6   &     645 \\
 & APM  &   239.5 & 764.4   & 421.3   &     366 \\
 & LBQS &    323.8 & 966.8   & 554.1   &     571 \\
\enddata
\tablerefs{APM = SMIH96; LBQS = WLFC95}
\end{deluxetable}
%\end{document}

%\clearpage
%  t8_lris.tex
%
%%%  fn gamma function fit results 
%
%%%%%%%%%%%%%%%%%%%%%
%\documentstyle[aas2pp4]{article}
%\tighten
%\begin{document}
%%%%%%%%%%%%%%%%%%%%%
\begin{deluxetable}{clll}
\tablewidth{0pc}
%\tabletypesize{\scriptsize}
%\scriptsize
\tablecaption{Fit to HI Column Density Distribution\label{t8}} 
\tablenum{8}
\tablehead{
\multicolumn{4}{c}{f(N,z)=( f$_*$ / N$_*$ ) ( N / N$_*$ )$^{-\beta}$ e$^{\rm -N/N_*}$} \\
\colhead{Redshift Range} &\colhead{f$_*$}& \colhead{log N$_*$} & \colhead{$\beta$} 
}
\startdata
\multicolumn{4}{c}{Including Lyman-limit System Data} \\
1.5 - 4.0 & 0.051 & $21.22^{+0.10}_{-0.16}$&1.24$^{+0.02}_{-0.08}$ \\ 
4.0 - 4.7 & 0.160 & $21.22^{+0.52}_{-0.33}$&1.08$^{+0.14}_{-0.22}$\\ 
\multicolumn{4}{c}{} \\
\multicolumn{4}{c}{Not Including Lyman-limit System Data} \\
1.5 - 4.0 & 0.03 & $21.39\pm0.30$ &$1.52\pm0.25$ \\
4.0 - 4.7 & 0.34 & $20.92\pm0.50$ &$0.32\pm0.60$ \\
\enddata
\end{deluxetable}
%%%%%%%%%%%%%%%%%%%%%
%\end{document}
%%%%%%%%%%%%%%%%%%%%%

%\clearpage
%  t9_lris.tex
%
%%%%%%%%%%%%%%%%%%%%%
%\documentstyle[aas2pp4]{article}
%\tighten
%\begin{document}
%%%%%%%%%%%%%%%%%%%%%
\begin{deluxetable}{rrrrr|rrr|rrr|rrr}
%\tabletypesize{\scriptsize}
\scriptsize
\tablewidth{0pt}
\tablecaption{Data for Figure 14\label{t9}}
\tablenum{9}
\tablehead{
\colhead{Redshift} &\colhead{DLA} & \colhead{$\Delta$ z}& \multicolumn{2}{c}{Number of} & 
\multicolumn{3}{c}{$\Omega = 1, \Lambda = 0$} &
\multicolumn{3}{c}{$\Omega = 0.3, \Lambda = 0$} &
\multicolumn{3}{c}{$\Omega = 0.3, \Lambda = 0.7$} \\
\colhead{Range} &  \colhead{$\langle{\rm z}\rangle$} & & \colhead{DLAs} & \colhead{QSOs} &
\colhead{$\Delta X$} & \colhead{$\Omega_g$} & \colhead{$\sigma$} &
\colhead{$\Delta X$} & \colhead{$\Omega_g$} & \colhead{$\sigma$} &
\colhead{$\Delta X$} & \colhead{$\Omega_g$} & \colhead{$\sigma$} 
}
\startdata
.008 -- 1.50 & 0.65 & 47.8 &  4 & 178 &   58.5 &0.54 &0.31 &   67.1 &0.47 &0.27&  80.7 &0.39 &0.22 \\
1.50 -- 2.00 & 1.84 & 96.4 & 19 & 372 &  162.1 &1.27 &0.36 &  219.1 &0.94 &0.27& 281.8 &0.73 &0.21 \\
2.00 -- 2.50 & 2.26 &124.2 & 25 & 349 &  223.0 &1.67 &0.56 &  310.3 &1.20 &0.40& 393.7 &0.94 &0.32 \\
2.50 -- 3.00 & 2.69 & 76.9 & 15 & 234 &  148.4 &1.54 &0.51 &  212.6 &1.08 &0.36& 265.1 &0.86 &0.29  \\
3.00 -- 3.50 & 3.21 & 40.9 & 10 & 128 &   83.9 &2.10 &0.80 &  123.0 &1.44 &0.54& 151.0 &1.17 &0.44 \\
3.50 -- 4.70 & 3.98 & 33.8 & 12 &  60 &   74.5 &1.23 &0.43 &  111.8 &0.82 &0.28& 134.6 &0.68 &0.24 \\
\enddata
\tablecomments{When totalled, the number of QSOs in column 5 is more than 
the total for this data set (646).  This is because the same QSO can contribute
redshift path to more than one redshift bin. }
\end{deluxetable}
%%%%%%%%%%%%%%%%%%%%%
%\end{document}
%%%%%%%%%%%%%%%%%%%%%

\end{document}